\documentclass[journal,onecolumn]{IEEEtran}

\pdfoutput=1

\usepackage{cite}
\usepackage{amsmath,amssymb,amsfonts}
\usepackage{algorithmic}
\usepackage{graphicx}
\usepackage{textcomp}
\usepackage[font=small,tableposition=top]{caption}
\usepackage[font=footnotesize]{subcaption}
\usepackage{xcolor}

\begin{document}
% \history{Date of current version October, 2018.}
% %\doi{10.1109/ACCESS.2017.DOI}

\title{The Multi-Scale Impact of the Alzheimer's Disease in the Topology Diversity of Astrocytes Molecular Communications Nanonetworks}
% \author{\uppercase{Michael Taynnan Barros}\authorrefmark{1}, ,
% \uppercase{Walisson Silva\authorrefmark{2}, and Carlos Danilo Miranda Regis}.\authorrefmark{2},
% \IEEEmembership{Member, IEEE}}
% \address[1]{Telecommunication Software \& Systems Group (TSSG), Waterford Institute of Technology, Ireland.}
% \address[2]{Federal Institute of Education, Science and Technology of Para\'iba, Brazil.}
% \tfootnote{The work described in this paper is facilitated by the Irish Research Council, under the government of Ireland post-doc fellowship (grant GOIPD/2016/650).}

% \corresp{Corresponding author: Michael Taynnan Barros (e-mail: mbarros@tssg.org).}

\author{Michael Taynnan Barros,
        Walisson Silva,
        and~Carlos Danilo Miranda Regis,% <-this % stops a space
\thanks{M. T. Barros is with the Telecommunication Software \& Systems Group (TSSG), Waterford Institute of Technology, Ireland. e-mail: mbarros@tssg.org (see https://michaeltaob.wixsite.com/mtaob).}% <-this % stops a space
\thanks{W. Silva. and C. D. M. Régis are with Federal Institute of Education, Science and Technology of Para\'iba, Brazil.}% <-this % stops a space
\thanks{The work described in this paper is facilitated by the Irish Research Council, under the government of Ireland post-doc fellowship (grant GOIPD/2016/650).}
\thanks{Manuscript submitted to a journal publication in October 2018.}}

\maketitle

% \fcolorbox{black}{grey}{\begin{minipage}{20em}
% Quality Checker
%     \begin{itemize}
%     \item[$\square$] Number of equations: $\leq 40$
%     \item[$\square$] Number of Figures for results: $\leq 20$
%     \item[$\square$] Number pages: $\leq 12$
%     % \item[$\boxtimes$] A closed item.
%     \end{itemize}
% \end{minipage}
% }

\begin{abstract}
The Internet of Bio-Nano-Things is a new paradigm that can bring novel remotely controlled actuation and sensing techniques inside the human body. Towards precise bionano sensing techniques in the brain, we investigate the challenges of modelling spatial distribution of astrocyte networks in developing a mathematical framework that lay the groundwork for future early-detection techniques of neurodegenerative disease. In this paper, we investigate the effect of the $\beta$-amyloid plaques in astrocytes with the Alzheimer's disease. We developed a computation model of healthy and Alzheimer's diseases astrocytes networks from the state of the art models and results that account for the intracellular pathways, IP$_3$ dynamics, gap junctions, voltage-gated calcium channels and astrocytes volumes. We also implemented different types of astrocytes network topologies including shortcut networks, regular degree networks, Erd\"os R\'enyi networks and link radius networks. A proposed multi-scale stochastic computational model captures the relationship between the intracellular and intercellular scales. Lastly, we designed and evaluated a single-hop communication system with frequency modulation using metrics such as propagation extend, molecular delay and channel gain.
The results show that the more unstable but at the same time lower level oscillations of Alzheimer's astrocyte networks can create a multi-scale effect on communication between astrocytes with increased molecular delay and lower channel gain compared to healthy astrocytes, with an elevated impact on Erd\"os R\'enyi networks and link radius networks topologies.
\end{abstract}

% \begin{keywords}
% Molecular communications, nanonetworks, bionano sensing, communication theory, Alzheimer's
% \end{keywords}

\section{Introduction}

\IEEEPARstart{M}{olecular} bionano devices is an exciting new technology that enables future integration of biological systems with the internet, termed the Internet of Bio-Nano-Things, towards minimally invasive high-resolution cellular interfaces \cite{Akyildiz:2015:ieeecommag}. Implantable bionano devices will then allow the remote real-time monitoring and treatment of diseases \cite{Barros2017,wirdatmadja2017wireless}. The molecular mechanisms that regulate any biological system homoeostasis are essential to be completely understood within a mathematical framework, which leads to a complete description of the signal propagation inside a biological communication network. The recent area of molecular communication has focused their efforts in the past decade in providing this framework for the analysis of communication system in living tissues even with different molecular information carriers, for example, DNA, Ca$^{2+}$, hormones, or even blood cells \cite{akan2017fundamentals}. However, there has been very little efforts in understanding these systems when they are found with a disease or any mechanism that lead to such.

Neurodegenerative diseases, as a whole, affect hundreds of millions of people around the world with an annual treatment cost of billions of dollars, and in particular the case of the Alzheimer's disease (AD), there is need of constant nurse care due to detrimental life symptoms \cite{haughey2003alzheimer}. Both the genesis and the cure of Alzheimer's are far away from being completely established, plus the complexity of this challenging research issue has attracted multidisciplinary methods in the past recent years. Now, research in Alzheimer's disease achieved increased progress by integrating neuroscience with computer science, nanotechnology, synthetic biology and communication engineering areas. Most recently, researchers are concentrating in two exciting new approaches to AD: i) trying to understand how AD affects the loss of neuronal information in the brain, and ii) developing cell-free disease models that potentially allow the progress of drug discovery research and testing with accurate computational models. 
% Organ-in-chip devices are compromised to the elimination of animal testing for drug-related research from the cell-free technology, where these multidisciplinary disease computational models have a fundamental role in their functionality.

The molecular information propagation inside the brain, in particular, involves different types of cells (neuronal and non-neuronal), which comprise different types of communication systems altogether. The organisation of astrocytes within a variety of networks is now being studied towards the detailed characterisation of the propagation of information signals\cite{Lallouette:2014:compneuroscience}. Neurodegeneration will also affect the signal propagation since they directly interfere with the multi-scale brain organisation, including the intracellular signal pathways, intercellular communications and higher hierarchical network topologies. Since the molecular communication signal propagation of neurodegenerative diseases is an interesting approach to understanding neurodegeneration at the cellular level, mixing its intracellular, intercellular and network modelling is a step further to more detailed information of its physiology using a computational multi-scale approach. Molecular communication analysis of neurodegenerative disease can unlock discoveries by augmenting the existing models for intracellular activity towards the quantification of higher brain levels.

\begin{figure}[!t]
    \centering
    \includegraphics[scale=0.5]{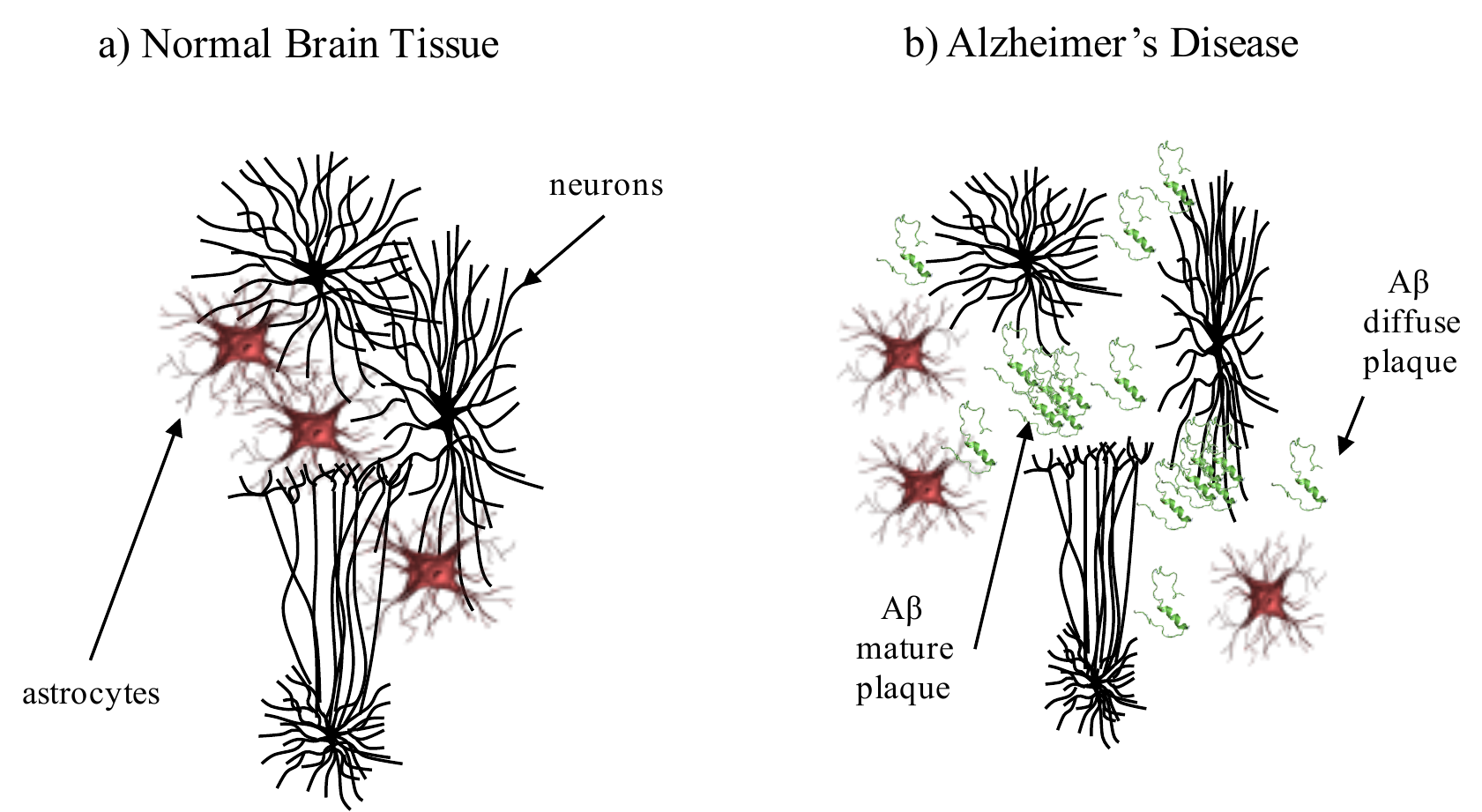}
    \caption{The effect of Amyloid-$\beta$ plaques in the communication between astrocytes and neurons, and its comparison to a healthy tissue. In the healthy tissue, the cells are well connected, which maintains the propagation of molecules and spikes inside a cell population. In the AD, the Amyloid-$\beta$ plaques, either diffuse or mature, brake the connections and block the propagation of signals. The Amyloid-$\beta$ plaques also are absorbed by the cells, changing their signalling pathways and also their morphology \cite{dal2018astrocytes}. We explore this latter issue here in this paper.}
    \label{fig:ad_effect}
\end{figure}

In AD, the intracellular calcium information propagation inside non-neuronal cells like astrocytes, impact on the tripartite synapses propagation in neuronal cells causing loss of connection, what is known in neuroscience as the synapse plasticity. The amyloid hypothesis suggests that the accumulation of $\beta$-amyloid in the brain is the primary driving force of AD pathogenesis and change of the synapses behaviour~\cite{hardy1991amyloid,selkoe2016amyloid,hardy2002amyloid,selkoe1991molecular}, \ref{fig:ad_effect}. In this hypothesis, the formation of $\beta$-amyloid (which is a self-aggregating 40-42 amino acid protein) plaques and fibrils as a result of an underlying imbalance process cause a slow accumulation these peptides that can alter calcium signalling processes leading to synaptic failure and neuronal death. Although the linkage between $\beta$-amyloid and intracellular calcium homeostasis remains unclear, several studies provide growing pieces of evidences that $\beta$-amyloid directly interferes with the inositol triphosphate (IP$_3$) \cite{demuro2013cytotoxicity,matrosov2017emergence}, the gap Junctions \cite{cruz2010astrocytic,haughey2003alzheimer}, the voltage-gated channels with the cellular membrane \cite{zeng:2009:biophysical}, and the cellular volume \cite{olabarria2010concomitant}. 

In this paper, our main goal is to quantify the effects of the intracellular calcium signalling signal propagation under $\beta$-amyloid influence, as a way to compute the AD affect inside astrocyte nanonetworks. We use the results of experiments and computational biology approaches to make simplifying assumptions for how $\beta$-amyloid affects the intracellular calcium signalling by creating a new computational model that accounts for the changes in calcium pathways, IP$_3$ dynamics, gap junctions, cellular membrane influence and cellular volume. Then, we also present a molecular simulation model that accounts for both intracellular and intercellular calcium signalling, that is a multi-scale method towards the quantification of the signal propagation inside astrocyte networks. The astrocyte 3D networks are modelled to account for topology diversity in the brain with shortcut networks, regular degree networks, Erd\"os R\'enyi networks and link radius networks. Lastly, to properly measure the effects of multi-scale communications we designed a calcium signalling-based molecular communication system in order to link the effects of the AD, compared to healthy networks, to signal propagation extend, molecular delay and channel gain. This throughout new computational approach is one step further to more accurate and complete models for AD research as well as providing novel insights about the molecular communication characterisation of diseases inside biological tissues, such as astrocytes.

Our main contribution are:
\begin{enumerate}
    \item \textbf{A computational Alzheimer's disease model based on the influence of $\beta$-amyloid plaques on the calcium cellular pathways.} The model accounts for calcium pathways, IP$_3$ dynamics, gap junctions, cellular membrane influence and cellular volume.
    \item \textbf{Implementation of different astrocyte network topologies}, including shortcut networks, regular degree networks, Erd\"os R\'enyi networks and link radius networks
    \item \textbf{Quantification of the multi-scale effects of the AD from an intracellular level, intercellular level and network level.} Our results show that the more unstable but at the same time lower level oscillations of Alzheimer's astrocyte networks can create a multi-scale effect on communication between astrocytes with increased molecular delay and lower channel gain compared to healthy astrocytes, with an increased impact in Erd\"os R\'enyi networks and link radius networks topologies.
\end{enumerate}

% The paper is organised as follows: 

\section{Related Work} \label{sec:related_work}

The research about calcium signalling in astrocytes is a recent attempt to depict the laws that govern the propagation of information in the brain, with the relationship between those cells and neurons \cite{giaume2010astroglial}. Many researchers have, through the years, provided the mathematical basis for the calcium signalling in astrocytes, including \cite{goldberg2010nonlinear,fujii2017astrocyte,matrosov2017emergence,manninen2017reproducibility,lenk2016understanding,wallach2014glutamate,szabo2017extensive}. Because there is an intrinsic dual relationship with neurons, astrocytes can change the synaptic plasticity dramatically of the neurons. Their network organisation, thereafter, has a fundamental impact in delivering information to the neurons encoded in many different molecules that regulate both their energy and functioning \cite{giaume2010astroglial}. Experimental pieces of evidence allow the observation of the propagation of calcium waves inside astrocyte networks, dependable on their brain region, but commonly exhibiting topology dependency in the information propagation dynamics \cite{Lallouette:2014:compneuroscience}. For example, this phenomenon is described in work about the olfactory glomeruli \cite{roux2011plasticity} or the somatosensory cortex \cite{houades2008gap}. Computational models are sufficient to accurately describe the same calcium wave propagation from the experiments with the topology variation \cite{Lallouette:2014:compneuroscience}. However, the usage of the experimental approaches is often limited to observe the full impact of the calcium signal propagation in these astrocyte network topologies. The intracellular results and output models are not used in terms of network functioning and changes. Molecular computational models can be introduced to quantify the signal propagation variability and the resulting models can account for intercellular communication properties of the astrocytes, i.e. gap junctions, network organisation and cellular population activity understanding.

Alzheimer's research has historically been concentrated on experimental approaches that analysed the genesis of the disease, and the ultimate result is the $\beta$-amyloid hypothesis. There are other peptides that are also correlated with the genesis of the AD, including the $\tau$ hypothesis. Researchers do not discard the possibility of these peptides working together, but the whole research on the genesis of the AD is slow due to the numerous efforts in the experimental validation of these hypotheses. In particular to astrocytes, the most advanced stages of understanding of AD and their calcium signalling are distributed in works that account for inositol triphosphate (IP3) \cite{demuro2013cytotoxicity,matrosov2017emergence}, the gap Junctions \cite{cruz2010astrocytic,haughey2003alzheimer}, the voltage-gated channels with the cellular membrane \cite{zeng:2009:biophysical}, and the cellular volume \cite{olabarria2010concomitant}.  More recently, researchers have started to create computational models to help the researcher understand how AD affects the brain in many scales and with different types of cells \cite{latulippe2018mathematical,toivari2011effects,de2013progression}. However, there is no effort towards compiling the recent experimental advancements from the literature in a novel computational model, moreover, there is no effort in analysing how the defect calcium signalling under the influence of AD from a molecular computational approach $\beta$-amyloid can affect the whole astrocytes networks.

From previous literature, we can infer that the amount of modelling in AD research is very limited. This open spaces for new molecular computational models that translate research results from practice to theory in order to enhance the disease analysis and contribute to its understanding from a computational approach. We can observe from the literature that the experimental investigation of the molecular principles of the AD is limited to obtain its overall understanding, for example, it is not possible to visualise the relationship of astrocyte network molecular communication and the spatial organisation of such network \cite{Pitta:2009:jbiophys}. Molecular computational models are well accepted in calcium signalling biology research area with a plurality of models that are validated through experiments, including astrocyte-to-astrocyte variability of intracellular signalling parameters, local IP$_3$ regeneration, receptor sub-types, affinity of IP$_3$ receptor-channels on calcium stores or kinetics of IP$_3$ transport through gap-junctions \cite{hofer2002control,Lallouette:2014:compneuroscience,iacobas2006stochastic}.

Astrocytes were also studied in the field of molecular communications with the dual objective of understanding both their natural communication properties and to transmit artificial information. Barros et. al. \cite{Barros:2015:tcom} compared how the astrocytes natural communication performance difference against other cell types including epithelial and smooth muscle cells. Mesiti et. al. \cite{mesiti2015astrocyte} took a different approach by studying how the astrocytes communicate to neurons in the tripartite synapses. More recently, Barros and Dey have developed a control of the calcium in the astrocytes cytosol upon pre-defined cellular stimulation \cite{barros:2017:ieeenano}, which enabled the development of effective communication system design that allows great performance in the transmission of bits in short-range communications \cite{barros2017feed}. The application of the above studies to biology and medicine through new bionanosensing techniques (see \cite{Barros:2014:Transnanoscience}) can create a great interdisciplinary impact, however, there are still major modelling and system design issues. One of them is the configuration of astrocyte 3D networks topologies that compose tissues. In \cite{Pitta:2009:jbiophys}, there are several astrocyte network topologies that were not taken into consideration in the molecular communication literature. This type of approach is also important to other types of cells, which their non-linear shapes and heterogeneous configuration can lead to a great impact on the communication system performance. On top of that, calcium signalling-based molecular communication system under disease scenarios is also a missing approach in reaching a full theoretical framework. We explore these gaps in the literature by modelling the astrocyte network topologies in a calcium-signalling based molecular communication system of the astrocytes, thus improving the existing efforts in understanding this complex communication channel. We also extend this model to account for the presence of AD in these communications systems, with the goal of analysing the signal propagation inside astrocytes networks when this type of neurodegeneration is at later stages of development.

\section{Ca$^{2+}$ signalling-based Molecular Communication System} \label{sec:calciumsignallingmodel}

\subsection{Signalling Model}

We model the calcium-signalling process in astrocytes based on the conventional and well accepted ODE equations by Lavrentovich and
Hemkin~\cite{Lavrentovich:2008:jtheobio}. The model is in accordance with experimental
observation \cite{Lavrentovich:2008:jtheobio}, and has been studied also in molecular communication systems by \cite{Barros:2015:tcom}. The three pool model
includes: Ca$^{2+}$ concentration in the cytosol ($C_a$)
(Eq. \ref{eq:xa}); Ca$^{2+}$ concentration in the endoplasmic
reticulum ($E_a$) (Eq. \ref{eq:ya}); and IP$_3$ concentration
($I_a$) (Eq. \ref{eq:za}). They are represented by the following
equations:
\begin{equation}\label{eq:xa}
\frac{dC_a}{dt} = \sigma_{0} - \kappa_{o}C_a + \sigma_{1} - \sigma_{2} + \kappa_f (E_a-C_a)
\end{equation}
\begin{equation}\label{eq:ya}
\frac{dE_a}{dt} = \sigma_{2} - \sigma_{1} - \kappa_f (E_a-C_a)
\end{equation}
\begin{equation}\label{eq:za}
\frac{dI_a}{dt} = \sigma_{3} -\kappa_{d}I_a
\end{equation}
\noindent where $\sigma_{0}$ is the flow of calcium from the extracellular space into the cytosol, $\kappa_{o}C_a$ is the rate of Ca$^{2+}$ efflux from the cytosol to the extracellular space, $\kappa_f (E_a-C_a)$ is the leak flux from the endoplasmic reticulum into the cytosol and $\kappa_{d}I_a$ is the degradation of IP$_3$. These terms and given by

\begin{eqnarray}\label{eq:v3astro}
\sigma_{1} & = & 4\Sigma_{M3}\frac{\kappa^{n}_{C1} C_a^n}{(C_a^n+\kappa^{n}_{C1})(C_a^n+\kappa^{n}_{C2})}\nonumber\\
& & .\frac{I_a^m}{\kappa^m_I + I_a^m}(E_a-C_a),
\end{eqnarray}
\begin{equation}\label{eq:vserca}
\sigma_{2} = \Sigma_{M2}\frac{C_a^2}{\kappa^2_2 + C_a^2},
\end{equation}
\begin{equation}\label{eq:vplc}
\sigma_{3} = \Sigma_p\frac{C_a^2}{\kappa^2_p + C_a^2},
\end{equation}

\noindent where the $\sigma_{1}$ term (Eq. \ref{eq:v3astro}), models the calcium flux from the endoplasmic reticulum to the cytosol via IP$_3$ stimulation. It is represented as, where $\Sigma_{m3}$ is the maximum flux value of Ca$^{2+}$ into the cytosol, $\kappa^{n}_{C1}$ and $\kappa^{n}_{C2}$ are the activating and inhibiting variables for the IP$_3$ %type-2 isoform, 
and the $m$ and $n$ are the Hill coefficients. The efflux of calcium from the sarco(endo)plasmic reticulum to the endoplasmic reticulum is modelled as $\sigma_{2}$, where $\Sigma_{M2}$ is the maximum flux of Ca$^{2+}$ in this process. Finally, $\sigma_{3}$ describes IP$_3$ generation by the Phosphoinositide phospholipase C (PLC) protein, where $\Sigma_p$ is the maximum flux of Ca$^{2+}$ in this process, and $p$ is the Hill coefficient.

\begin{figure}
    \centering
    \includegraphics[scale=0.35]{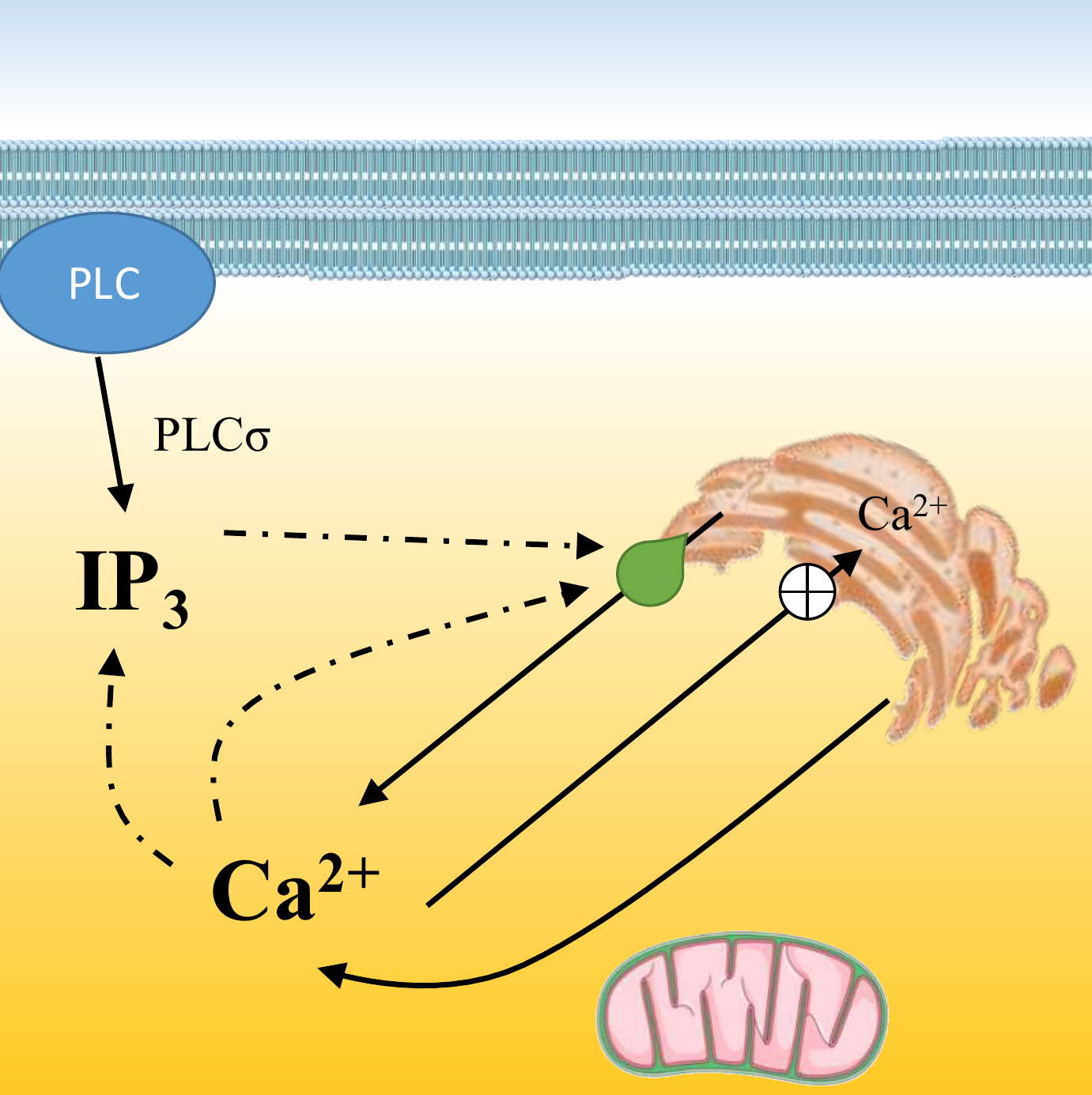}
    \caption{The healthy astrocyte calcium (Ca$^{2+}$) model. Both the relationship between the calcium and IP$_3$ is regulated by the PLC protein and the endoplasmic reticulum.}
    \label{fig:calcium_model}
\end{figure}

\subsection{Gap Junction Model} \label{sec:gap_junction_model}

The gap junction stochastic model introduced by Baigent et al. \cite{Baigent:1997:jtheobio} has been first studied for molecular
communication by Kilinc and Akan \cite{Kilinc:2013:nanotech} and later on by Barros et. al. \cite{Barros:2015:tcom}. We
consider voltage-sensitive gap junctions with two states of
conductance for each connexin: an open state with high conductance and a closed state with low conductance. Thereafter, the four basic combinations of states are considered: \textit{State HH}: Both gates are in a high conductance state. This probability is denoted by $p_{HH}$; \textit{State HL}: One gate is in a high conductance state and the other is in a low conductance state. This probability is denoted by $p_{HL}$; \textit{State LH}: One gate is in a low conductance state and the other is in a high conductance state. This probability is denoted by $p_{LH}$; \textit{State LL}: Both gates are in a low conductance state. This probability is denoted by $p_{LL}$. Experimental validation of the model indicated that the $LL$ state appears to present very low occurrence rates \cite{Bukaukas:2013:biophycs}, thus we neglect that state here. Thus, the probabilities should follow $p_{HH} + p_{HL} + p_{LH} = 1.$

Moreover, $p_{HH}$, $p_{HL}$ and $p_{LH}$ are interrelated as follows:
\begin{equation}\label{eq:phl}
\frac{dp_{HL}}{dt} = \beta_1 (\vartheta_j)\times p_{HH} - \alpha_1(\vartheta_j) \times p_{LH}
\end{equation}
\begin{equation}\label{eq:plh}
\frac{dp_{LH}}{dt} =  \beta_2 (\vartheta_j)\times p_{HH} - \alpha_2(\vartheta_j) \times p_{HL}
\end{equation}
\noindent where the control of the gap junctions is mediated through the potential difference of the membrane of two adjacent cells ($\vartheta_j$), the gate opening rate is $\alpha$ and gate closing rate is $\beta$. The terms $\alpha_1(\vartheta_j)$, $\alpha_2(\vartheta_j)$, $\beta_1(\vartheta_j)$ and $\beta_2(\vartheta_j)$ are defined as $\alpha_1(\vartheta_j) = \lambda e^{-A_\alpha (\vartheta_j - \vartheta_0)}$, $\alpha_2(\vartheta_j) = \lambda e^{A_\alpha (\vartheta_j + \vartheta_0)}$, $\beta_1(\vartheta_j) = \lambda e^{A_\beta (\vartheta_j - \vartheta_0)}$, $\beta_2(\vartheta_j) = \lambda e^{-A_\beta (\vartheta_j + \vartheta_0)}$, where $\vartheta_0$ is the junctional voltage at which the opening and closing rates of the gap junctions have the same common value $\lambda$, and $A_\alpha$ and $A_\beta$ are constants that indicate the sensitivity of a gap junction to the junctional voltage.

\section{Astrocyte calcium signalling model under the influence of Alzheimer's disease} \label{sec:alzheimersmodel}

The $\beta$-amyloid hypothesis is essential to capture the changes in the calcium signalling in astrocytes with the goal of quantifying the signal propagation in the brain, as a whole. Within the later stages of the AD, the $\beta$-amyloid plaques in the brain will interact with the cells, and in astrocytes, the calcium concentration is changed overall \cite{berridge2016inositol}. The temporal and spatial effects on this increase are not well covered by the neuroscience community and the characterisation of the relationship between affected astrocytes and their networks is still missing. Since experimental efforts in this scenario are rare due to the infrastructural capabilities and resources limitations, computational approaches are the most accessible method to study such systems. Towards this direction, we present an astrocyte intracellular calcium-signalling model under the influence of $\beta$-amyloid plaques as an approach to studying the AD effect on these cells. Our model is an attempt to integrate both existing validated computational models and experimental results in the direction of a more complete description of the studied scenario. It is not a simple integration process since its construction was carefully carried out to maintain a valid relationship of linear and non-linear relationships intact between variables.

\subsection{Signalling Model}

We use a baseline model for the intracellular calcium-signalling of astrocytes based on the work of \cite{toivari2011effects}. We use four pool ODE to quantify the changes over time of: the cytoplasmic Ca$^{2+}$ ($C$), endoplasmic reticulum Ca$^{2+}$ concentration $S$, IP$_{3}$ concentration ($I$) and active IP3R (R). First, consider

\begin{figure}
    \centering
    \includegraphics[scale=0.35]{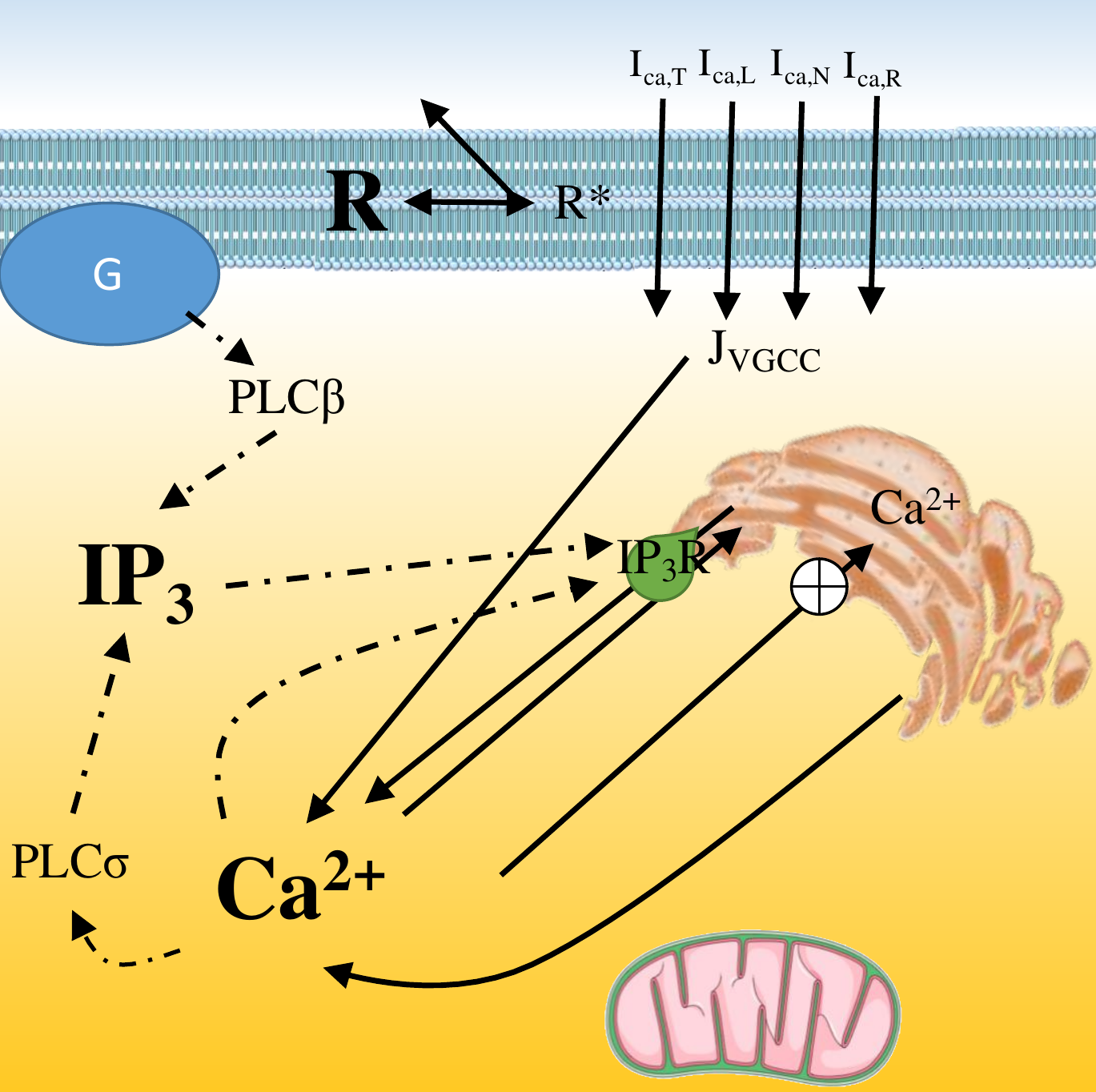}
    \caption{The astrocyte calcium (Ca$^{2+}$) model under $\beta$-amyloid influence (AD). Both the relationship between the calcium and IP$_3$ is regulated by the G protein, the PLC protein, the endoplasmic reticulum abd the voltage-gated cellular membrane.}
    \label{fig:calcium_model}
\end{figure}

\begin{equation} \label{eq:cad}
\begin{split}
 \frac{dC}{dt} = & J_{VGCC} + v_{CCE} + v_{iR} - V_{OUT}\\
 & + V_{ER(leak)} + V_{ER(rel)} - V_{SERCA}   
\end{split}
\end{equation}

\begin{equation}\label{eq:ead}
    \frac{dE}{dt} = \beta\, (v_{SERCA} - v_{ER(leak)} - v_{ER(rel)})
\end{equation}

\noindent where $J_{VGCC}$, $v_{CCE}$, $v_{iR}$, $V_{OUT}$, $V_{ER(leak)}$, $V_{ER(rel)}$ and $V_{SERCA}$ which represent respectively the voltage-gated calcium channel, CCE calcium uptake, the inotropic receptor rate, calcium extrusion, rate of calcium leakage from the endoplasmic reticulum, rate of release of calcium from the endoplasmic reticulum and calcium pumping into the endoplasmic reticulum. In Equations \ref{eq:cad} and \ref{eq:ead}, the effects of (fast) calcium buffering in the cytoplasm and in the endoplasmic reticulum stores are accounted for by defining effective rate constants and an effective diffusion coefficient.

We assume a non-linear rate of regulating capacitance Ca$^{2+}$ as described in \cite{di2007calcium}

\begin{equation}
    v_{CCE} = \frac{k_{CCE}H_{CCE}^2}{H_{CCE}^2 + E^2}
\end{equation}

\noindent where $k_{CCE}$ and $H_{CCE}$ are the maximal rate constant for CCE influx and the half-inactivation constant for CCE influx respectively. The rate of calcium influx, induced by ionotropic receptors, from ECM to cytosol was modelled as in \cite{di2007calcium}:

\begin{equation}
    v_{iR} = k_{ATP(P2X)} \frac{[ATP]_{ex}^{1.4}}{(H_{ATP(P2X)} + [ATP]_{ex}^{1.4})}
\end{equation}

\noindent where $k_{ATP(P2X)}$ and $H_{ATP(P2X)}$ are the maximal rate of stimuli-evoked inotropic calcium influx and the half-saturation constant for stimuli-evoked ionotropic calcium. 

Similarly, the activation of G protein and PLC$\beta$ pathways, induced by metabotropic receptors, to promote the IP$_3$ production were reformed from \cite{di2007calcium,hofer2002control} and modelled as:

\begin{equation}
    v_{PLC\beta} = k_{ATP(P2Y)} \frac{[ATP]_{ex}}{(K_D + [ATP]_{ex})}
\end{equation}

\noindent where $k_{ATP(P2Y)}$ and $K_D$ are the maximal rate of IP$_3$ production mediated by the metabotropic receptors and the dissociation constant for the binding of ligand/metabotropic receptors.

The following rate laws equation is chosen to model IP$_3$-induced and calcium-induced calcium release, as well as the endoplasmic reticulum calcium regulation are modelled in the following as:

\begin{equation}
    V_{OUT} = k_5 C
\end{equation}

\begin{equation}
    v_{ER(leak)} = k_1 (E - C)
\end{equation}

\begin{equation}
    v_{ER(rel)} = \frac{k_2 R C^2 I^2}{[(K_a^2 + C^2)(K_{IP3}^2 + I^2)]}(E - C)
\end{equation}

\begin{equation}
    v_{SERCA} = k_3 C
\end{equation}

\noindent where the $k_5$, $k_1$, $k_2$, $k_a$, $K_{IP3}$ and $k3$ are the rate of calcium extrusion from plasma membrane, the rate of calcium leak from the endoplasmic reticulum, the rate of calcium release through IP$_3$ receptors, the half-saturation constant for calcium activation of the IP$_3$ receptor and the half saturation constant for IP$_3$ activation of the corresponding receptor, respectively. Even though there are pieces of evidence that the endoplasmic reticulum suffers stress under the $\beta$-amyloid influence\cite{alberdi2013ca2+}, we have not found enough data to construct a model based on the Hill equations, therefore, we do not explore this phenomenon in detail in our model.

\subsection{IP$_3$ Dynamics}

The IP$_{3}$ dynamics of astrocytes is essential to model the signalling process of astrocytes with AD due to the direct control of the calcium mimicking the experimental results. In addition to the activation of the IP3R by cytoplasmic calcium, we also account for a slower calcium-induced inactivation We use the IP$_{3}$ concentration ($I$) and active IP3R (R), described as

\begin{equation}\label{eq:rad}
    \frac{dR}{dt} = v_{IP3R(Rec)} - v_{IP3R(Inact)}
\end{equation}

\begin{equation}\label{eq:iad}
    \frac{dI}{dt} = J_{prod} - v_{IP3(Deg)} + \mathcal{D}_i^{I}(I_i,I_j|j \in \mathcal{N}_i)
\end{equation}

\begin{equation}
    v_{IP3(Deg)} = k_9 I
\end{equation}

\begin{equation}
    v_{IP3R(Rec)} = k_6 \frac{K_i^2}{(K_i + C^2)}
\end{equation}

\begin{equation}
    v_{IP3R(Inact)} = k_6R
\end{equation}

\noindent where $k_9$, $k_6$, and $k_i$ are the rate constant of IP$_3$ degradation, the rate constant of IP$_3$ receptor inactivation and the half saturation constant for calcium activation of the IP$_3$ receptor, respectively.

The modelling of complex calcium signals have to include stochastic processes in the IP$_3$ pathway, and in this way, mimic the $\beta$-amyloid effects on the intracellular calcium \cite{matrosov2017emergence}. Adding a further term for stochastic IP$_3$ production either by spontaneous PLC$\delta$ activation or by activation of PLC$\beta$ by stochastic synaptic inputs \cite{Lavrentovich:2008:jtheobio,aguado2002neuronal}. Accordingly, IP$_3$ production by PLC isoenzymes is described by \cite{wu2014spatiotemporal}

\begin{equation} \label{eq:ip31}
\begin{split}
    J_{prod}(t)& = O_{\beta}\mathcal{U}_{\beta}(0,1|t_k)\sum_k \delta(t - t_k) \\
    &+ O_{\delta}\mathcal{U}_{\delta}(0,1|\upsilon_n)\sum_n \delta(t - \upsilon_n)
\end{split}
\end{equation}

\noindent where $\mathcal{U}_x(0, 1|t) (with x = \beta, \delta)$ denotes the generation of a random number at time $t$ following a uniform distribution, and represents random modulations of the maximal rate of IP$_3$ production; $t_k$ and $v_n$ respectively represent the instants of synaptically evoked and spontaneous nucleation of IP$_3$/calcium spikes, which can be assumed to be approximately Poisson distributed, following experimental observations \cite{softky1993highly,skupin2008does}.

To account for the sparse distribution of IP$_3$ inside the network, we incorporated its diffusion in the overall IP$_3$ dynamics model. The diffusion is defined as a summation of a set of $j$ flows from cell $i$ within the set of neighbouring cells $\mathcal{N}_i$, and defined as

\begin{equation} \label{eq:ip32}
    \mathcal{D}_i^{I}(I_i,I_j|j \in \mathcal{N}_i) = \sum_{j \in \mathcal{N}_i} J_{ij}
\end{equation}

\noindent each flow is defined as

\begin{equation} \label{eq:ip33}
\begin{split}
    J_{ij}& = -\frac{F_{ij}}{2} \left(1 + \tanh\left(\frac{|\Delta I_{ij} - I_{\theta}|}{\omega_I}\right)\right) \\
    &\frac{\Delta I_{ij}}{|\Delta I_{ij}|}
\end{split}
\end{equation}

\noindent where $\Delta I_{ij}$ is the IP$_3$ gradient between $j$ and $i$, $I_{\theta}$ is the diffusive IP$_3$ gradient threshold, $\omega_I$ increase slope of $\Delta I_{ij}$, and $F$ is the maximum diffusion coefficient flow. This model enables the direct multi-scale influence of $\beta$-amyloid by interfering with the production of IP$_3$ inside the astrocytes and its propagation in the network by the above presented non-linear transport process.

\subsection{Gap Junction Model}

We investigated how the existing efforts in the literature could impact the gap junction model presented in Section \ref{sec:gap_junction_model}. Experimental studies have presented results for gap junction behaviour under the influence of $\beta$-amyloid for astrocytes in both in-vitro and in-vivo settings \cite{cruz2010astrocytic}\cite{haughey2003alzheimer}. These studies observed that even with the changes in the calcium pathways, amplitude and signal velocity, the gap junctions performance were not changed and therefore inconclusive results were reached. Therefore, we use the same previously presented gap junction model in our proposed model.

\subsection{VGCC models of the cellular membrane}

Voltage-gated calcium channels are mechanisms that increase the calcium concentration of cells due to the membrane voltage, which can come from a different source, but in there, the different types is a wholesome approach to quantifying the overall values of the influence. Studies about the influence of $\beta$-amyloid in the VGCC of astrocytes can be found in \cite{zeng:2009:biophysical}. We want to calculate the current contribution to the increase of calcium concentration, the current was converted into the flux as

\begin{equation} \label{eq:vgccc1}
    J_{VGCC} = - \frac{I_{VGCC}}{zFV_{ast}}
\end{equation}

\begin{equation}\label{eq:vgccc2}
    I_{VGCC} = I_{Ca,T} + I_{Ca,L} + I_{Ca,N} + I_{Ca,R}
\end{equation}

\begin{table}\label{tab:vgcc}
    \centering
    \caption{Concrete formula for every typeof calcium currents}
    \begin{tabular}{ll}
        \hline
        \bf Channel type & \bf Equation of channel kinetics \\\hline
        \textbf{T-type} & $I_{Ca,T} = \overline{g}_T m_T (h_{Tf} + 0.04h_{Ts})(V - E_{Ca})    $\\
         & $\overline{m}_T = \frac{1.0}{1.0 + e^{-(V + 63.5)/1.5}}\overline{h}_T = \frac{1.0}{1.0 + e^{(V + 76.2)/3.0}}$\\
         & $\tau_{h_{Tf}} = 50.0 e^{-((V + 72)/10)^2} + 10$ \\
         & $\tau_{h_{Tf}} = 400.0 e^{-((V + 100)/10)^2} + 400$ \\
         & $\tau_{m_{T}} = 65.0 e^{-((V + 68)/6)^2} + 12$ \\
         \textbf{L-type} & $I_{Ca,L} = \overline{g}_L m_L h_L (V - E_{Ca})    $\\
         & $\overline{m}_L = \frac{1.0}{1.0 + e^{-(V + 50.0)/3.0}}\overline{h}_L = \frac{0.00045}{0.00045 + Ca_{cyt}}$\\
         & $\tau_{m_{L}} = 18.0 e^{-((V + 45)/20)^2} + 1.5$ \\
        \textbf{N-type} & $I_{Ca,N} = \overline{g}_N m_N h_{N}(V - E_{Ca})    $\\
        & $\overline{m}_N = \frac{1.0}{1.0 + e^{-(V + 45.0)/7.0}}$\\
        & $h_N = \frac{0.00010}{0.00010 + Ca_{cyt}}$\\
        & $\tau_{m_{N}} = 18.0 e^{-((V + 70)/25)^2} + 0.30$ \\
        \textbf{R-type} & $I_{Ca,R} = \overline{g}_R m_R h_{R}(V - E_{Ca})    $\\
        & $\overline{m}_R = \frac{1.0}{1.0 + e^{-(V + 10.0)/10.0}}\overline{h}_R = \frac{1.0}{1.0 + e^{(V + 48.0)/5.0}}$\\
        & $\tau_{m_{R}} = 0.1 e^{-((V + 62.0)/13.0)^2} + 0.05$ \\
        & $\tau_{h_{Tf}} = 0.5 e^{-((V + 55.6)/18.0)^2} + 0.5$ \\
    \end{tabular}
\end{table}

\noindent where $V_{ast}$ is the volume of an astrocyte assumed as a spherical soma with a radius of $5 mm$. $J_{VGCC}$ is the change rate of calcium concentration. And $I_{Ca,T}$, $I_{Ca,L}$, $I_{Ca,N}$ and $I_{Ca,R}$ the steady-state current of VGCCs that gives the total of calcium current flow through VGCCs in an
individual cell ($I_{VGCC}$). The concrete formula for every type of calcium current is given in detail in Table I.  In this way, not only the total calcium flux but also network influx via different VGCCs can be calculated.

\subsection{Astrocyte Volume Model} \label{sec:volume_model}

The authors \cite{olabarria2010concomitant} using an immunohistochemical approach that observed the atrophy of astrocytes by measuring the surface, volume, and the relationship between astrocytes and neuritic plaques. Their results are relevant to temporal observations of volume changes in astrocytes. Since in our presented study we concentrate on the later stages of the AD, we use their results to modify the volume of astrocytes from 19.635~$\mu$m$^3$ to 11.027~$\mu$m$^3$, representing a total of 43.84\% of the initial cell volume.

\section{Network Topologies Model} \label{sec:networktopologiesmodel}

The spatial distribution of astrocytes and their connectivity can dictate the spatial signal propagation pattern of Ca$^{2+}$ waves and change the molecular communication performance using this channel. It is important that is characterised and analysed so that new techniques can emerge to deal with this type of issue in the molecular communications community. Astrocytes networks are characterised to exhibit a large average number of connections
per cells and that contain long-distance connections, small values of the mean shortest path and long-distance connections could instead promote signal propagation. The mapping process of astrocyte network topologies is challenging and relies upon sophisticated experimental approaches that combine both the molecular process that regulates their signalling process and imaging approaches that obtain their topology set up. In \cite{Lallouette:2014:compneuroscience}, the authors presented a collection of five topologies that were collected from the diverse literature review in the area including regular degree, link radius, shortcut networks, and Erd\"os R\'enyi. They were able to connect astrocytes network structures to known graph theory models. We use the same methodology to model the astrocyte network topologies and furthermore integrate with our molecular computational model presented in the previous sections, which are detailed in the following. In the end, these models will enable us to obtain also the spatial-temporal characterisation of calcium signal propagation in astrocyte network topologies.

% \Figure[t!](topskip=0pt, botskip=0pt, midskip=0pt)[scale=0.6]{astnet}
% {Astrocyte Networks and their locations in the brain (MAKE IT REAL).\label{fig:ast_net}}

\begin{figure}
    \centering
    \includegraphics[scale=0.5]{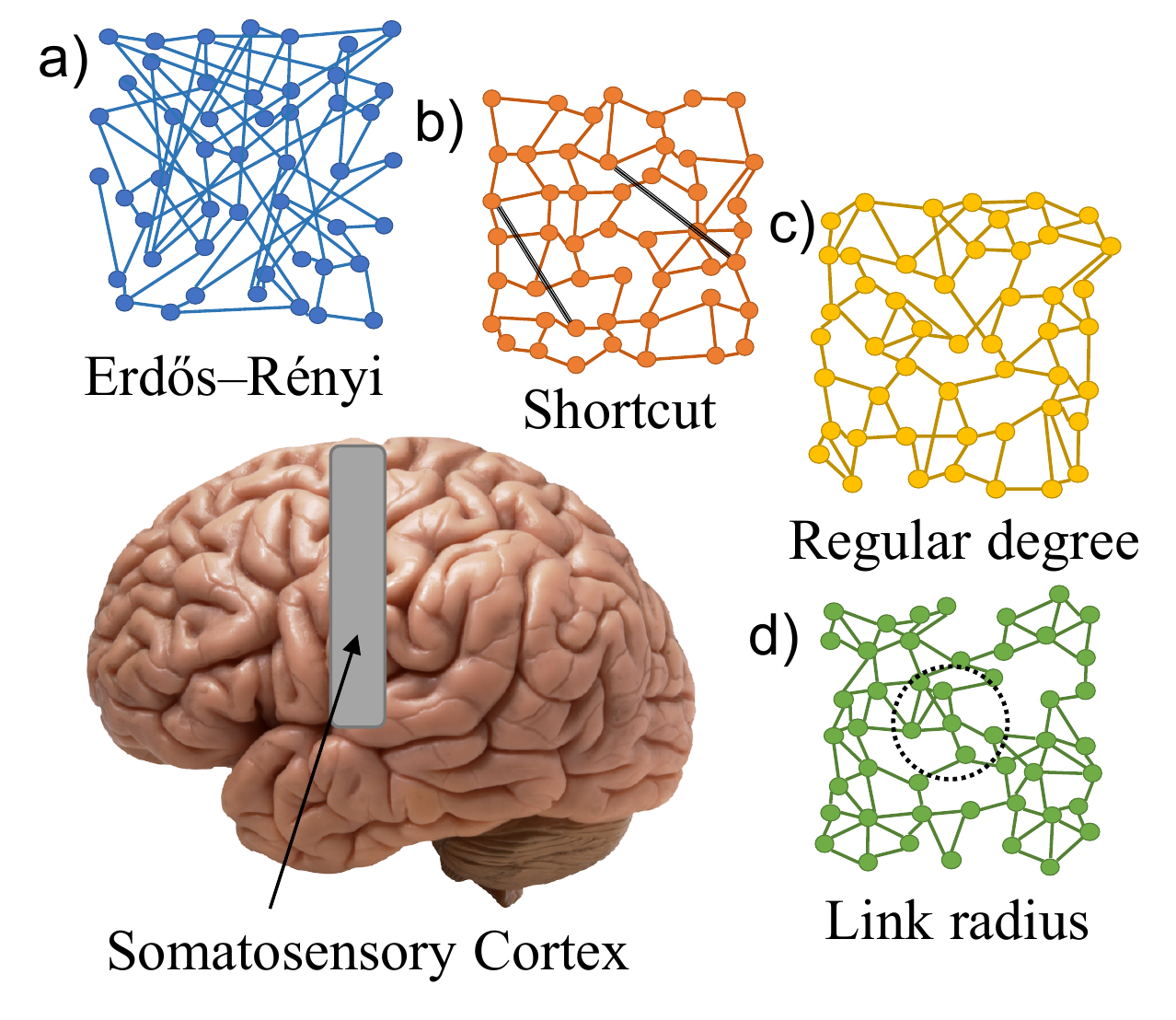}
    \caption{The different types of 3D Astrocyte Networks in the somatosensory cortex of the brain \cite{houades2008gap}.}
    \label{fig:ast_net}
\end{figure}

\begin{figure*}[!]
    \centering
    \includegraphics[scale=0.7]{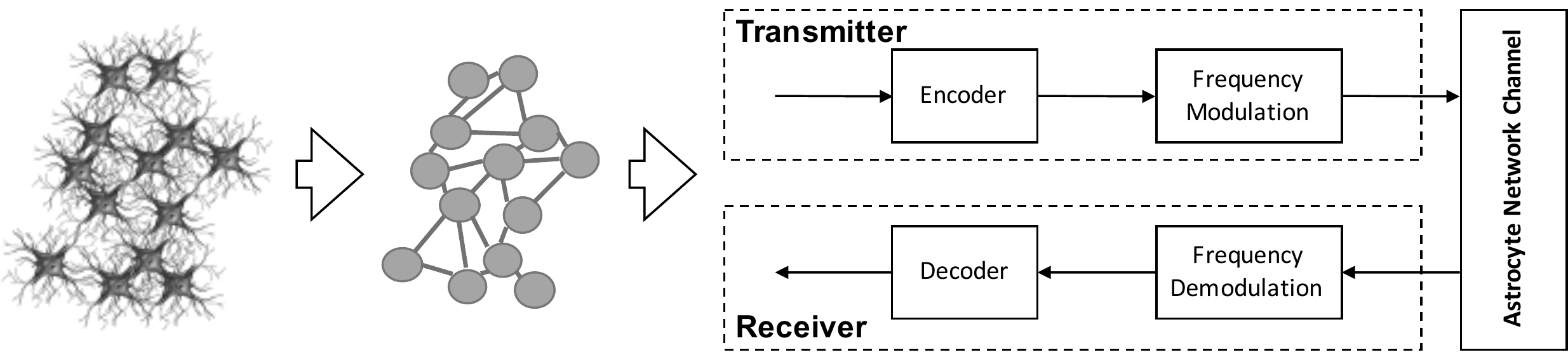}
    \caption{Model of the Astrocyte network Calcium-based Molecular Communication System. Real astrocyte networks are modelled as a graph and used to construct the connections between cells. This graph is used then with the computational models of single cell astrocyte to create the multi-scale stochastic simulations. Transmitter and receiver are also designed.}
    \label{fig:sys_model}
\end{figure*}

\subsection{Types and models}

We consider a cellular tissue space discretized into a 3D cubic lattice matrix ($S$) composed of $I \times J
\times K$ cells ($c$), where $c_{i,j,k}$ ($i=1 \dots I$; $j=1, \dots
J$ and $k=1, \dots K$) denotes an arbitrary cell in the tissue. The
cells are connected with a maximum of six neighbouring cells. The organisation
of the cells is assumed to be a layered 3D lattice. Defining the topologies inside the 3D lattice is based on a graph that represents the tissue on a network level. This separation allows us to provide a flexible way to account for both intracellular pathways and intercellular signal propagation inside a pre-defined topology, and construct the multi-scale model. We consider a graph $G$ composed with $c$ cells as nodes and the set $T$ of links (edges), which each element is defined as a tuple of two arbitrary cells $(c_{i,j,k},c_{i',j',k'})$ where ($i=1 \dots I$; $j=1, \dots
J$ and $k=1, \dots K$). The cells are connected to a fixed number of other cells, and therefore $|T| = n x |c|$, which $n$ is the number of connection per cell. In the following, we characterise the different topology models based on this initial description graph model of the spatial distribution of astrocytes within a given tissue area. We present the implementation details for each used topology:

\begin{enumerate}
    \item \texttt{Regular degree networks:} This type of network topology refers to $n$ as a static number for all cells, thus providing a homogeneous environment. Cells are connected to their nearest neighbours. 
    \item \texttt{Link radius networks:} Consider a random variable $z$ that $\in [0,n]$ and represents the number of connections per cell inside an Euclidean distance $d$. $d$ can be varied to increase the dynamism in the network topology organisation.
    \item \texttt{Shortcut networks:} This topology, defined as "small world" networks are used for structures dependable on randomly connected cells at the edges of the network. Within $S$ we introduce $a$ that is the internode distance, and then, connect to the nearest neighbours with are multiple of $a$ or $\leq a \times |S|$. In our implementation, this internode distance will connect directly the transmitter to a receiver in a communication channel. The remaining nodes of the network will follow the regular distribution with the $n$ number of neighbours.
    \item \texttt{Erd\"os R\'enyi networks:} is used for providing unconstrained spatial connection distribution between the nodes. We consider that cells are connected based on a probability $p$. In this way, each cell will connect to the maximum of six cells selected arbitrarily. This choice is only performed for the first neighbouring cell, and the other connection choices are kept with the $p$ probability, so we can avoid an increasingly complex network topology.
\end{enumerate}

% \subsubsection{Spatial scale-free networks} Astrocyte network topologies can also present a very large distribution of connections inside $S$, which can also be termed as "hub" astrocytes Xu et al., 2010. We used the attachment rule by placing sequentially astrocytes and, after every interaction, connecting new ones to the ones with the larger degree within spatial distance constraints. Consider a number of $m_{sf}$ astrocytes that represent the existing network topology. We couple a newcomer astrocyte to $m_{sf}$ with respect to maintaining the probability $Pi->j$ that the chosen $j$ astrocyte that increases the network with $j+1$ degrees and decreases its distance with pi→j ∝ kj exp(−dij/rc) where i is the index of
% the newcomer and dij the distance between i and j. The parameter
% rc controls the trade-off between scale–free structure and
% the restriction of the couplings to short distances (Barthélemy,
% 2010) {\bf DID WE IMPLEMENT IN THIS WAY}. 
% For large values of rc (larger than 20 μm in our case),
% some of the astrocytes (“hubs”) are coupled to a very large
% number of cells but the GJC couplings extend over large distances.
% On the contrary, small rc values (less than 5 μm in our
% case) lead to networks that are devoid of hubs but that feature
% short distance GJC couplings.
 %{\bf HOW DID WE IMPLEMENT THIS} 

\section{Communication System Analysis} \label{sec:comm_sys}

% \Figure[t!](topskip=0pt, botskip=0pt, midskip=0pt)[scale=0.4]{astcom}
% {Astrocyte Networks Communications (MAKE A NEW VERSION OF THIS).\label{fig:astcom}}

We designed a communication system that sends information through the astrocytes networks so we can measure the signal propagation effects in both scenarios and calculate the propagation extend, molecular delay and channel gain. The communication system is shown in Fig. \ref{fig:sys_model}, and consist of a transmitter, receiver and channel, which are depicted in Fig. \ref{fig:sys_model} and described in the following.

\begin{enumerate}
    \item \texttt{Transmitter:} The transmitter is a cell positioned in the centre of an intermediary cut of the cellular tissue. This astrocyte is assumed to have an infinite source of calcium based on a stimulation mechanism that allows the execution of signal frequency modulation.
    \item \texttt{Channel:} The channel is the 3D astrocyte network. We model the different topologies that an astrocyte network can form, which describe the organisation of connection between the cells and is further explained in Section \ref{sec:networktopologiesmodel}. Each astrocyte in the network has an initial calcium concentration that is independently controlled by their regulatory calcium pathways and also has their random intercellular signals propagation. This mechanism is important for the regeneration of calcium along the tissue but can become a source of noise or interference.
    \item \texttt{Receiver:} The receiver is an astrocyte that is a few cells away from the transmitter. Once the calcium arrives in a determined concentration, the receiver is activated and initiates the reception of the calcium molecules. For this analysis, we assume no intracellular interference and optimal reception ligand reception. 
\end{enumerate}
%\textbf{WALISSSON TO WRITE THIS PART}

\begin{table*}[t!] \label{tab:sim_variables}
    \centering
    \caption{Values of parameters to the Updated Model.}
    \begin{tabular}{lcccc}
        \hline
        Parameter & Reference & Normal Value & Alzheimer Value & Unit \\\hline
        $k_1$ & \cite{alberdi2013ca2+} & 0.0004 & 0.1 & 1/s \\
        $k_2$ & \cite{alberdi2013ca2+} & 0.2 & 1.5 & 1/s \\
        $k_3$ & \cite{alberdi2013ca2+} & 0.5 & 0.1 & 1/s \\
        $k_5$ & \cite{alberdi2013ca2+} & 0.5 & 0.05 & 1/s\\
        $k_6$ & \cite{alberdi2013ca2+} & 4 & 0.5 & 1/s\\
        $k_9$ & \cite{alberdi2013ca2+} & 0.08 & 0.5 & 1/s\\
        $v_7$ & \cite{alberdi2013ca2+} & 0.02 & 15 & $\mu$M/s\\
        $K_{IP3}$ & \cite{alberdi2013ca2+} & 0.3 & 0.5 & $\mu$M\\
        $K_a$ & \cite{alberdi2013ca2+} & 0.2 & 2.02 & $\mu$M \\
        $K_i$ & \cite{alberdi2013ca2+} & 0.2 & 0.15 & $\mu$M \\
        $K_{Ca}$ & \cite{alberdi2013ca2+} & 0.3 & 0.15 & $\mu$M \\
        $\beta$ & \cite{alberdi2013ca2+} & 35 & 0.1 & -- \\
        $H_{CCE}$ & \cite{alberdi2013ca2+} & 10 & 40 & $\mu$M\\
        $k_{CCE}$ & \cite{alberdi2013ca2+} & 0.01 & 2.2 & $\mu$M/s \\
        $k_{ATP(P2X)}$ & \cite{alberdi2013ca2+} & 0.08 & free & $\mu M$/s \\
        $H_{ATP(P2X)}$ & \cite{alberdi2013ca2+} & 0.09 & \textbf{0.15} & $\mu$M\\
        $k_{ATP(P2Y)}$ & \cite{alberdi2013ca2+} & 0.5 & \textbf{0.5} & $\mu$M/s\\
        $K_D$ & \cite{alberdi2013ca2+} & 10 & 0.8 & $\mu$M\\
        $z$ & \cite{zeng2009simulation} & 2 & -- \\
        $T$ & \cite{zeng2009simulation} & 300 &  & K \\
        $R$ & \cite{zeng2009simulation} & 8.31 &  & J/(mole$\cdot$ K) \\
        $F$ & \cite{zeng2009simulation} & 8.31 &  & J/(mole$\cdot$ K) \\
        $R$ & \cite{zeng2009simulation} & 8.31 &  & J/(mole$\cdot$ K) \\
        $F$ & \cite{zeng2009simulation} & 96,485 &  & J/(Coul/mole) \\
        $\overline{g}_T$ & \cite{zeng2009simulation} & 0.06 &  & pS \\
        $\overline{g}_L$ & \cite{zeng2009simulation} & 3.5 &  & pS \\
        $\overline{g}_N$ & \cite{zeng2009simulation} & 0.39 &  & pS \\
        $\overline{g}_R$ & \cite{zeng2009simulation} & 0.2225 &  & pS \\
        $V_{ast}$ & \cite{zeng2009simulation} & 5.223 $\times$ 10$^{-13}$ &  & l \\
        $O_{\beta}$ & \cite{matrosov2017emergence} & 0.15 &  & $\mu$M/s \\
        $O_{\delta}$ & \cite{matrosov2017emergence} & 0.15 &  & $\mu$M/s \\
        $t_{k}$ & \cite{matrosov2017emergence} & 0.3 &  & s \\
        $v_{n}$ & \cite{matrosov2017emergence} & 30 &  & s \\
        $F$ & \cite{lallouette2014modelisation} & 3.64 &  & s$^{-1}$ \\
        $I_{\theta}$ & \cite{lallouette2014modelisation} & 0.15 &  & $\mu$M \\
        $\omega_{I}$ & \cite{lallouette2014modelisation} & 0.05 &  & $\mu$M \\
    \end{tabular}
\end{table*}

\subsection{Multi-scale Molecular Computational Model} \label{sec:multiscalesim}

We use the Gillespie stochastic direct method (\cite{gillespie:1997:phychems}) as a mean to produce accurate variability of the chemical reactions for the models described in Sections \ref{sec:calciumsignallingmodel} and \ref{sec:alzheimersmodel}.  Consider the same cellular space introduced in Section \ref{sec:networktopologiesmodel}. Using this process, at each time step, the Gillespie algorithm is executed to select a random cell and a random internal reaction of that cell, also scheduling a time step ($t$) to each one of them.
Now, consider that each cell contains a set of internal reactions of the pools obtained from the previous reaction-diffusion models. This
stochastic solver computes the values of each pool over time,
selecting and executing scheduled reactions. The pool will be negatively
or positively affected by a constant $\alpha$ when a certain reaction is
executed. 

%In Sections \ref{sec:excitable_channel}, \ref{sec:non-excitable_channel} and \ref{sec:hybrid_channel}, we presented different models describing Ca$^{2+}$ behavior using reaction-diffusion equations. 

The process of executing one of the distinct reactions in $R$ requires a scheduling process divided into two phases---selecting a reaction and selecting a time step. Each reaction is allocated a reaction constant ($a_r$). % where the $a_r$ value is exactly the value of variable that models the reaction {\bf ** don't understand this equation}. 
Considering that $\alpha_0$ is the summation of all $a_r$ in $R$, the next reaction chosen $r_u$ will be:
\begin{equation}
r_u = \mbox{MAX}\left\{\frac{a_{r_j}}{\alpha_0} = \frac{a_{r_j}}{\sum\limits_{j=1}^{|R|} a_{r_j}}\right\}, u \in \mathbb{N}, u \in R
\end{equation}
\noindent which follows the \textit{roulette wheel selection} process, which selects the events based on their probability values. However, $u$ must satisfy the following restriction:
\begin{equation}
\sum\limits_{j=1}^{u-1}  \frac{\alpha_{r_j}}{\alpha_0}< \rho_2 \leq \sum\limits_{j=1}^{u} \frac{\alpha_{r_j}}{\alpha_0}
\end{equation}
\noindent in which $\rho_2$ is a uniform random variable with values in the range $(0,1)$.

At each time step ($t$), a time lapse ($\tau_t$) is derived based on $\alpha_0$, and is represented as:
\begin{equation}
\alpha_0 \cdot \tau_t = \mbox{ln} \frac{1}{\rho_1}
\end{equation}
\noindent in which $\rho_1$ is a uniform random variable with values in the range $(0,1)$. This process ends when $\sum\limits_{i=0}^{|T|} \tau_i < t_\theta$, where $T$ is the set of $t$ and $t_\theta$ is the maximum simulation time.

% The effect of executing a reaction is basically changing the values of the pools. According to the differential equation in hand, a constant will change the value of the pool according to the positive or negative effect of the executed reaction. The probability of selecting a reaction ($P(r_j,\tau_t)$) is defined as follows:
% \begin{equation}\label{eq:prob_cell}
%     P(r_j,\tau_t) dt = \alpha_{r_j} e^{- \alpha_0 \tau_t} dt
% \end{equation}

% The $P(r_j,\tau_t)$ favours a cell $c_{i,j,k}$ with a high value in its pool.
% Consequently, a high quantity of Ca$^{2+}$ will lead to diffusion, which will model the propagation of the molecules across the cellular space.

%The amount of diffusion reactions will be $N_r$, where $N_r = 3S \times N_g$ and $N_g$ is the number of GJs in each possible diffusion direction.

%Now consider a set of $N$ $\times$ $S$ distinct reactions, defined as $R$ with $|R| = C \times S$ {\bf **I would take this last sentence out cause it doesn't seem like we use it}.

We extend the Fick's diffusion model to account for the propagation of molecules with gap junctions in a cellular tissue area. In this way, we are able to capture the temporal-spatial dynamics of intercellular Ca$^{2+}$ signalling. We use Ca$^{2+}$ concentration difference to model this temporal-spatial characteristic, as follows \cite{Nakano:2010:Ieeenanobio}:
\begin{equation}\label{eq:diffusion_gjs}
Z_{\Delta} (i,j,k,n,m,l) = \frac{D}{v}(|Z_{n,m,l} - Z_{i,j,k}|) \times p_{(.)}
\end{equation}
\noindent where $n$ $\in$ $(i-1,i+1)$, $m$ $\in$ $(j-1,j+1)$, $l$ $\in$
$(k-1,k+1)$, $D$ is the diffusion coefficient, $v$ is the volume of the
cell, and $Z_{\Delta}$ is the difference in Ca$^{2+}$ concentration
between the cells. $p_{(.)}$ is the probability of the gap junction
opening and closing. Based on the gap junction probabilities, we define 
three different diffusion reactions for each cellular connection. Such reactions
are the multiplication of the probabilities ($p_{HH}$, $p_{HL}$ and $p_{LH}$) with the
regular intercellular diffusion probability. Our simulations suggest
that changing the organisation of gap junction connections is enough
to reproduce the large variability of the range of calcium signal propagation observed in experiments. For more information about the modelling consider the paper \cite{Barros:2015:tcom}.

\begin{figure*}
    \centering
    % \begin{subfigure}[b]{0.3\linewidth}
    %     \includegraphics[width=\textwidth]{prop_extent}
    %     \caption{}
    % \end{subfigure}
    ~ %add desired spacing between images, e. g. ~, \quad, \qquad, \hfill etc. 
    \begin{subfigure}[b]{0.32\linewidth}
        \includegraphics[width=\textwidth]{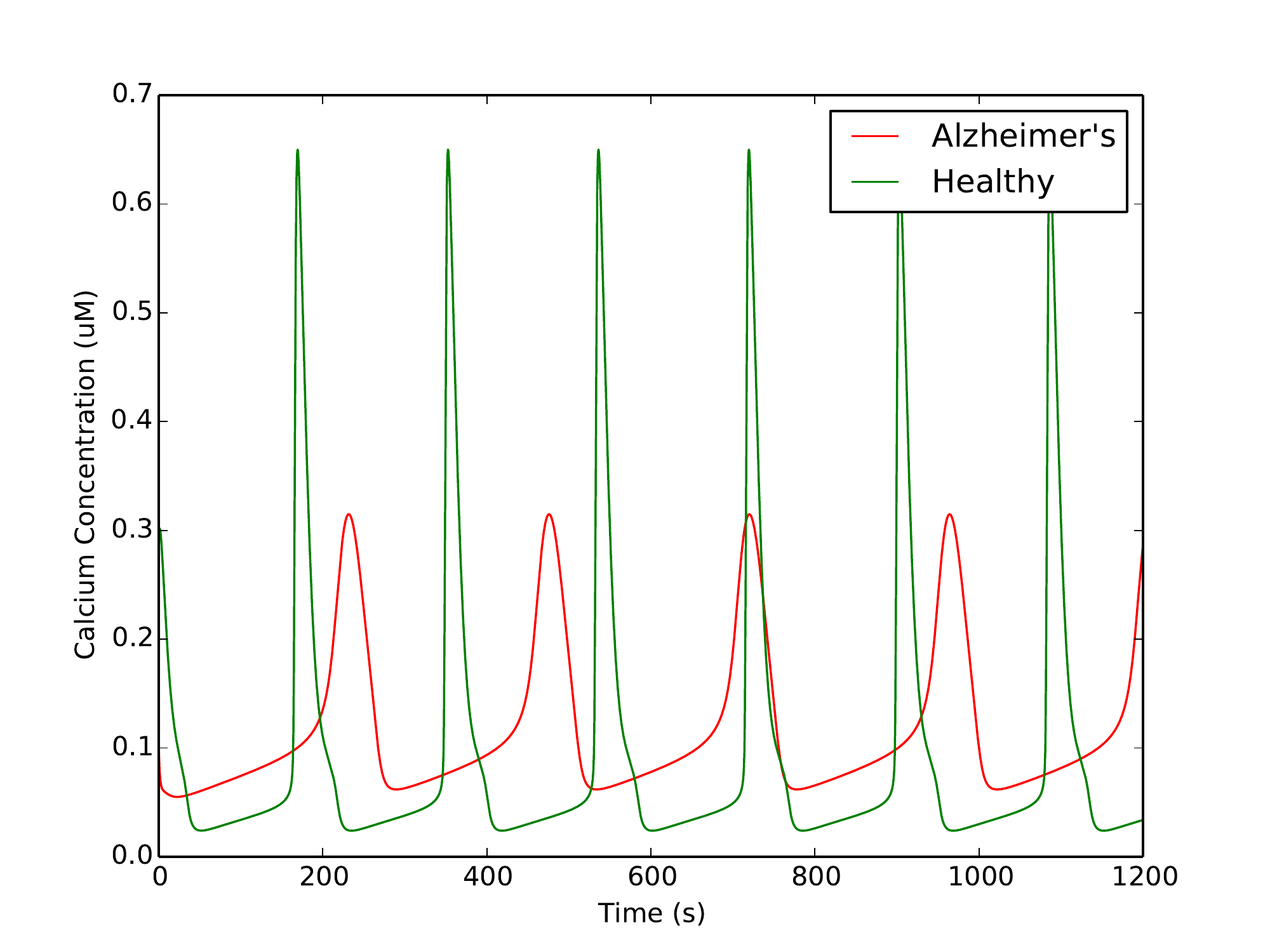}
        \caption{}
    \end{subfigure}
    \begin{subfigure}[b]{0.32\linewidth}
        \includegraphics[width=\textwidth]{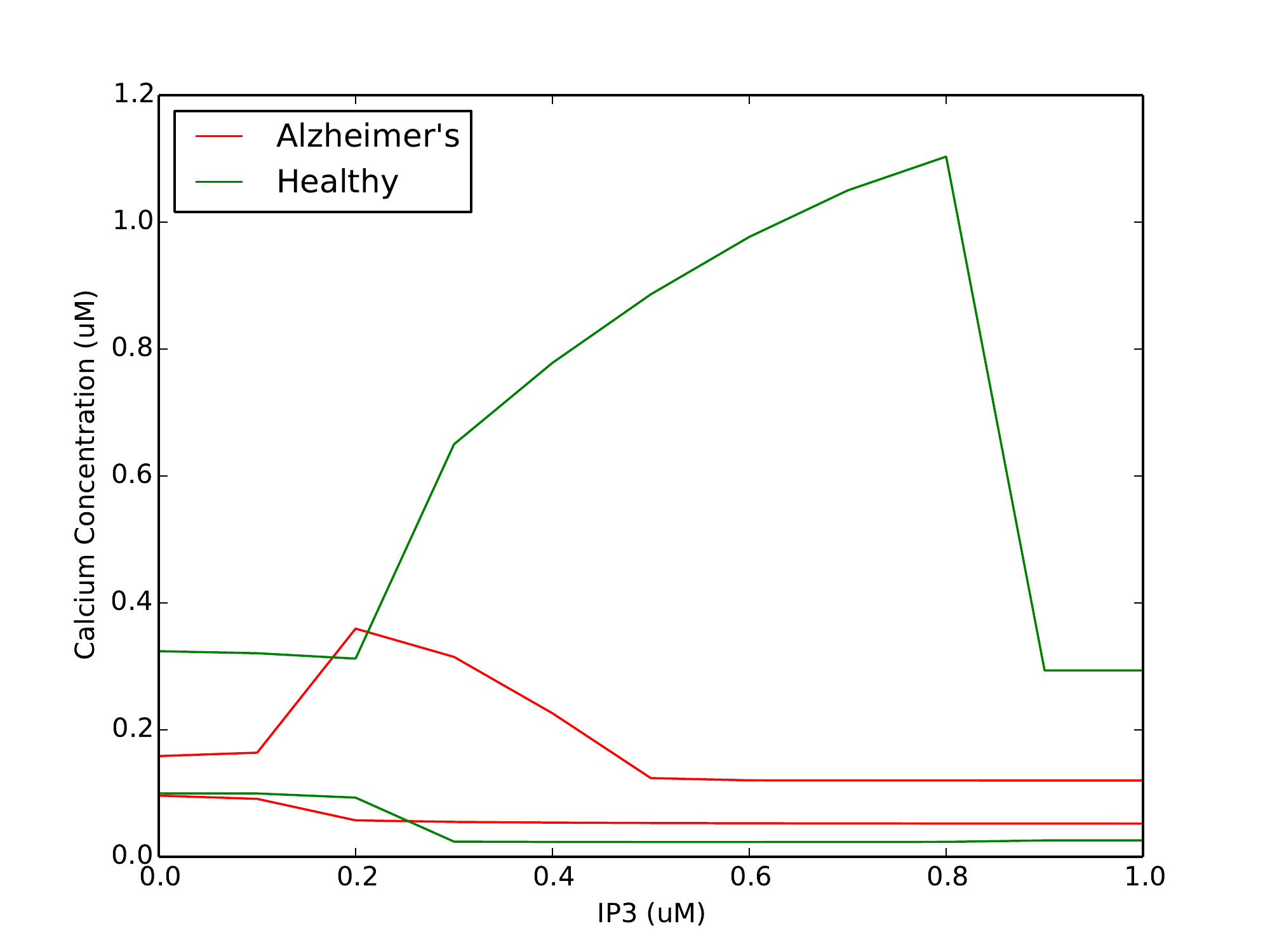}
        \caption{}
    \end{subfigure}
    \begin{subfigure}[b]{0.32\linewidth}
        \includegraphics[width=\textwidth]{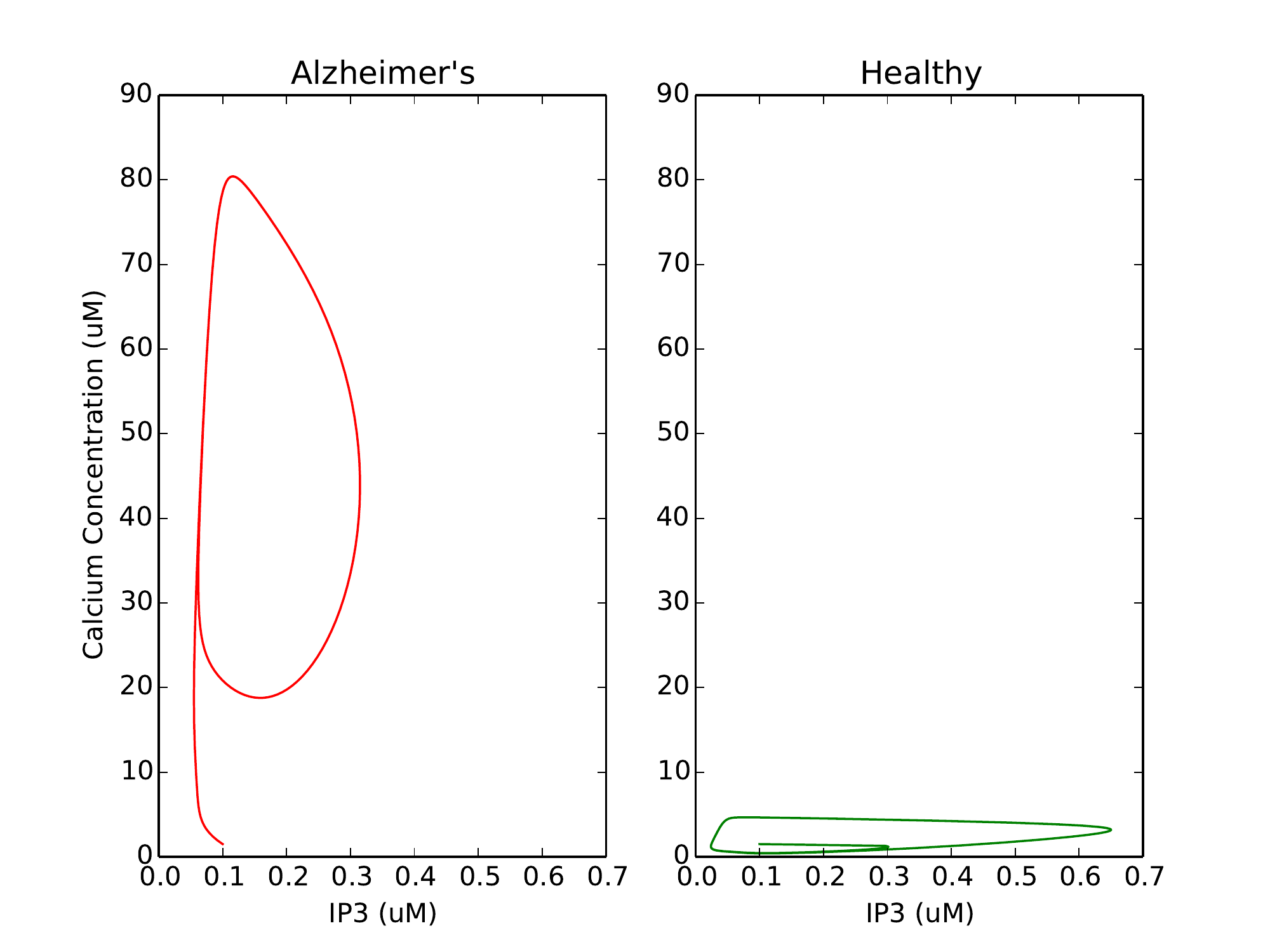}
        \caption{}
    \end{subfigure}
    \caption{(a) Calcium oscillation patterns for healthy and Alzheimer's astrocytes with fixed IP$_3$ increasing coefficient of 0.3 (b) Maximum and minimum calcium oscillation peaks for healthy and Alzheimer's astrocytes (c) Bifurcation analysis of calcium concentration and IP$_3$ for healthy and Alzheimer's astrocytes}
    \label{fig:single_cell} 
\end{figure*}

\subsection{$\beta$-Amyloid Assumptions}

Due to the unknown effects of $\beta$-Amyloid on the overall organisation and function of astrocytes networks and their calcium signalling \cite{latulippe2018mathematical}, we are obligated to consider a few assumptions:\\
\noindent \texttt{Assumption 1:} There are different development stages of AD that are correlated with the different concentration of $\beta$-Amyloid.  We consider only a short timescale within the studied stage of the AD. It is not our objective to model changes of $\beta$-Amyloid over time.\\
\noindent \texttt{Assumption 2:} We consider that the AD is at later development stages and that the $\beta$-Amyloid is sparsely distributed with a homogeneous effect on the astrocytes network. There is currently no detailed spatial description of $\beta$-Amyloid diffusion inside the brain.\\
\noindent \texttt{Assumption 3:} The networks are maintained at constant configuration throughout the whole analysis. This neglects the effects of cellular proliferation and cellular death that may come from the cellular stress caused by a certain concentration of $\beta$-Amyloid, however, due to \textit{Assumption 1}, this assumption does not hugely impact on our results.

\section{Results} \label{sec:results}

Our main objective is to obtain a detailed description of the changes that the presence of the $\beta$-Amyloid has on the signal propagation of calcium within the molecular communication system embedded into an astrocyte network. These changes are quantitatively aggravated when there is topology diversity meaning, we will use the presented models to obtain also the changes of signal propagation inside that scenario. We used the models presented in Sections \ref{sec:calciumsignallingmodel}, \ref{sec:alzheimersmodel}, \ref{sec:networktopologiesmodel} and \ref{sec:comm_sys}. First, we analyse the single cell results for a single-scale analysis, and thereafter, we present results for the multi-scale analysis using metrics such as \textit{propagation extent}, \textit{molecular gain} and \textit{molecular delay} for performance evaluation. The values for the simulation variables can be found in Table II. This whole computation model allows us to get the correlation of multiple scales of the astrocytes communication processes, starting from the intracellular scale (0-15$\mu$m) till the intercellular signalling and networking (15-800$\mu$m) and provide a comparison between healthy and AD astrocytes. We declare that the comparison between the healthy tissue scenario \ref{sec:calciumsignallingmodel} and the AD scenario \ref{sec:alzheimersmodel} are valid for the communication system analysis but they are not sufficient to drive conclusions about treatment strategies for the disease just yet.

\subsection{Single-scale analysis}

In Fig. \ref{fig:single_cell}, we show the results in solving the differential equations for healthy tissues (Eq. \ref{eq:xa},\ref{eq:ya} and\ref{eq:za}) and AD (Eq. \ref{eq:cad}, \ref{eq:ead}, \ref{eq:rad} and \ref{eq:iad}). This will provide sufficient understanding of the effects of $\beta$-amyloid in astrocytes before we present the multi-scale analysis. In Fig \ref{fig:single_cell} (a), we present the calcium oscillation patterns for healthy and AD scenarios. Both the frequency and peak of the oscillation change dramatically, which is explained by the overall effects of the $\beta$-amyloid changes in the cell including IP$_3$ dynamics, VGCC channels and intracellular pathways. The same changes are better captured by the calcium concentration response to the variation of IP$_3$ in maximum and minimum values, in Fig. \ref{fig:single_cell} (b). In the AD, the oscillations are confined to lower difference in the values that calcium can reach. However, In Fig. \ref{fig:single_cell} (c), we show by the bifurcation analysis that even with lower differences of calcium values, in the AD, the signal in the cytosol is unstable.

\begin{figure}
    \centering
    \begin{subfigure}[b]{0.4\linewidth}
        \includegraphics[width=\textwidth]{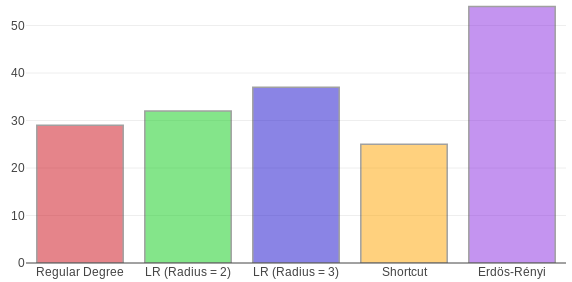}
        \caption{}
    \end{subfigure}
    ~ %add desired spacing between images, e. g. ~, \quad, \qquad, \hfill etc. 
    \begin{subfigure}[b]{0.4\linewidth}
        \includegraphics[width=\textwidth]{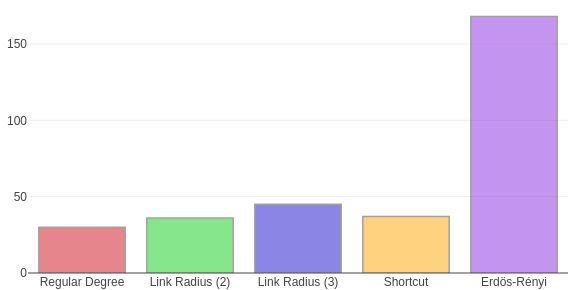}
        \caption{}
    \end{subfigure}
    \caption{(a) Propagation extent to each topology for the healthy scenario. (b) Propagation extent to each topology for AD.}
    \label{fig:prop_extent} 
\end{figure}

\begin{figure*}
    \centering
    % \begin{subfigure}[b]{0.3\linewidth}
    %     \includegraphics[width=\textwidth]{prop_extent}
    %     \caption{}
    % \end{subfigure}
    ~ %add desired spacing between images, e. g. ~, \quad, \qquad, \hfill etc. 
    \begin{subfigure}[b]{0.3\linewidth}
        \includegraphics[width=\textwidth]{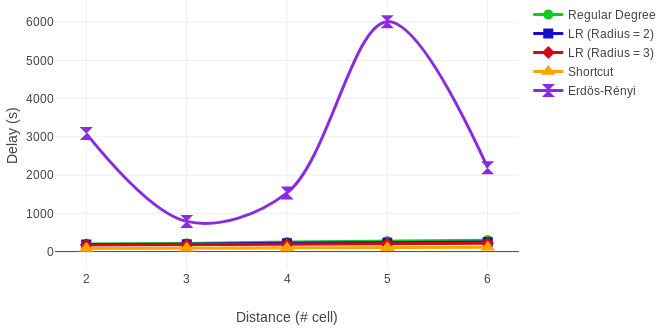}
        \caption{}
    \end{subfigure}
    \begin{subfigure}[b]{0.3\linewidth}
        \includegraphics[width=\textwidth]{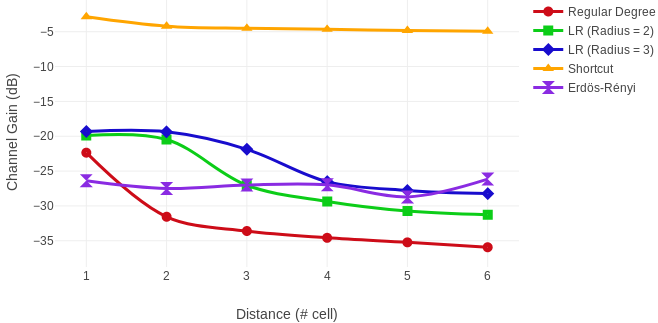}
        \caption{}
    \end{subfigure}
    \begin{subfigure}[b]{0.3\linewidth}
        \includegraphics[width=\textwidth]{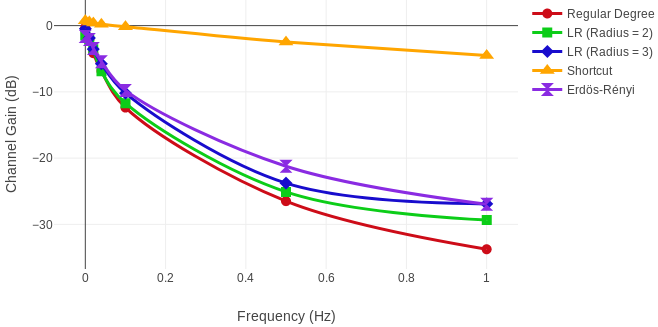}
        \caption{}
    \end{subfigure}
    \caption{Communication system metrics results for the healthy tissue scenario. (a) Molecular delay as a function of the distance (number of cells between transmitter and receiver). (b) Channel gain as a function of the distance (number of cells between transmitter and receiver). (c) Channel gain as a function of the calcium oscillations.}
    \label{fig:norm} 
\end{figure*}

\begin{figure*}
    \centering
    % \begin{subfigure}[b]{0.3\linewidth}
    %     \includegraphics[width=\textwidth]{prop_extent}
    %     \caption{}
    % \end{subfigure}
    ~ %add desired spacing between images, e. g. ~, \quad, \qquad, \hfill etc. 
    \begin{subfigure}[b]{0.3\linewidth}
        \includegraphics[width=\textwidth]{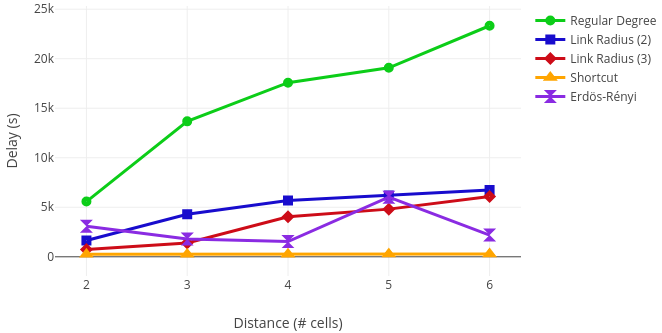}
        \caption{}
    \end{subfigure}
    \begin{subfigure}[b]{0.3\linewidth}
        \includegraphics[width=\textwidth]{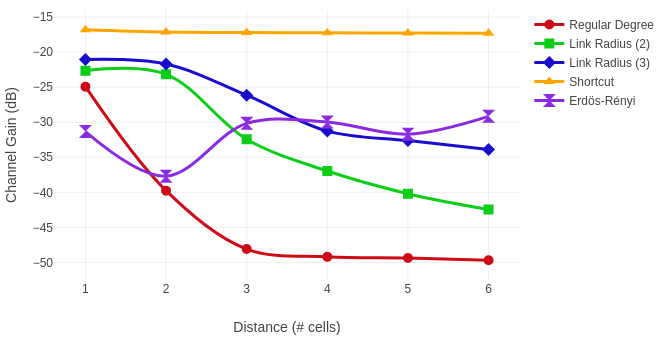}
        \caption{}
    \end{subfigure}
    \begin{subfigure}[b]{0.3\linewidth}
        \includegraphics[width=\textwidth]{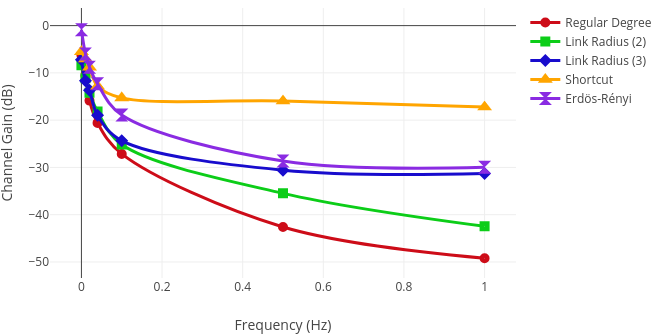}
        \caption{}
    \end{subfigure}
    \caption{Communication system metrics results for the AD tissue scenario. (a) Molecular delay as a function of the distance (number of cells between transmitter and receiver). (b) Channel gain as a function of the distance (number of cells between transmitter and receiver). (c) Channel gain as a function of the calcium oscillations.}
    \label{fig:alzh} 
\end{figure*}

% \subsection{Spatio-Temporal Ca$^{2+}$ Concentration Dynamics}

\subsection{Multi-scale analysis}

In the presented multi-scale analysis we move from a single cell scenario to capture the effects of calcium propagation in both health and AD scenario in the scale of intercellular communication and network topologies. The stochastic simulator presented in Section \ref{sec:multiscalesim} allows this phenomenon to be captured by a computational approach.

\noindent \texttt{1) Propagation Extent:} The propagation extent of calcium signals in the scenarios we are studying can be found in Fig. \ref{fig:prop_extent}. In this metric, we analysed how the transmission of one astrocyte will trigger the activation of calcium production in a number of different cells. Our results suggest that the level of network organisation can hugely impact on the in the number of activated cells, which means, that depending on the distance between the transmitter and receiver, this can convert in interference from other cells. Especially for Erd\"os R\'enyi networks, where the dynamics of generating connections can ultimately increase the overall propagation extent. For both healthy and AD astrocyte networks, the performance of Erd\"os R\'enyi is superior to the remaining topologies. While the other topologies have a propagation extent around 30 for healthy tissues and 25 for AD tissues, Erd\"os R\'enyi topologies present 50 for healthy tissues and 150 for AD tissues. Another interesting information from these results is that the effect of $\beta$-amyloid in the signal propagation can either increase and decrease the propagation extent. This means that there is an optimum topology arrangement that is aggravating the calcium propagation in different ways, and since the topologies resemble different parts of the brain, it is correlated in how the calcium signals can either provide more or fewer cells impacted by the presence of AD. 

\noindent \texttt{2) Molecular Delay:} The molecular delay is a time-dependent measure of a concentration of calcium from the transmitter to the receiver \cite{Barros:2015:tcom}, and depicted in Fig \ref{fig:norm} (a) and Fig. \ref{fig:alzh} for both healthy and AD astrocytes.  First, it is worth mentioning the different patterns of delay that are obtained from the different network topologies for both healthy and AD. The Erd\"os R\'enyi network topology is presented with an oscillatory delay dependent on distance variation due to its connection probability distribution over the network, which can create more paths to the receiver nodes and allowing more calcium to flow towards them. The same level effect is not observed in the other topologies as delays increases over distance. There is a delay increase in the AD, which its factor is dependent on the network topology. The regular degree is found to have the highest delay increase factor moving from around 100s of delay to a maximum of 25000s. The increase for the other types of the network was more modest, coming from an average around 2000s to a maximum of 5000s. This increase can be explained by the alteration of both cellular volume and VGCC channels, which are changed by the effect of $\beta$-amyloid in astrocytes. By the decrease of cellular volume (\ref{sec:volume_model}), the diffusion flow rate is changed \label{eq:diffusion_gjs} accordingly, and this increased effect will propagate the molecules quicker and thus increasing the multi-paths propagation inside the tissue. For a network with the regular degree, that means the homogeneous configuration of the network contributes to this dispersion of molecules causing the multi-paths effects. The VGCC channels will contribute to this effect since the post-synaptic voltage stimulus is captured by Eqs. \ref{eq:vgccc1} and \ref{eq:vgccc2}, and the decrease of neural rhythms is a direct modulator of astrocytes activity.

\noindent \texttt{3) Channel Gain:} The channel gain is a measure of how much of the molecules initial sent have been received within a given pre-determined interval. We analysed the channel gain for both distance between transmitter and receiver (\ref{fig:norm} (b) and Fig. \ref{fig:alzh} (b)) and frequency of oscillations in the transmitter (\ref{fig:norm} (c) and Fig. \ref{fig:alzh} (c)). For all graphs, it is clear that the shortcut topology presents the best channel gain performance, which is expected with an average of $-10$dB. For increased randomness of the network connection probabilities, for topologies Link Radius (3) and Erd\"os R\'enyi, the channel capacity performs better, depending on the distance and the frequency. Link Radius (3) has a more homogeneous distribution that allows the short distance cells to receive more molecules with an average of $-22$dB, however, for longer distances and also increase frequency, the heterogeneous configuration of the Erd\"os R\'enyi allows an increased reception of calcium by the increased multi-path propagation with an average of $-28$dB. Because this is subject to time settings, both increasing or decreasing the pre-determined interval might change the presented results. Lastly, the AD seems to worsen the calcium propagation in the cellular tissue due to the IP$_3$ dynamics that create non-linear effects of the calcium regeneration mechanism inside the cells from Eq. \ref{eq:ip31}, \ref{eq:ip32} and \ref{eq:ip33}.

% \subsection{Fading analysis}

% From the results of the frequency response find an appropriate fading distribution and a regression formula that fits the distribution

% \subsection{Shortest path analysis}

% Shortest path of network topologies model

% modify cost functions of nodes with a normalized ca2+ value 

% recalculate it per time and signal power versus frequency?

\section{Discussion}

The multi-scale effect in the brain is an essential mechanism to capture the signal propagation of different types of signals in the brain since the intracellular relationship of brain cells can affect the intercellular communication, the networks of cells and ultimately the networks of networks \cite{avena2018communication}. This complexity is poorly understood by neuroscientists, and there are very few proposed physical bionano interfaces that enable the study of this phenomenon in-vivo \cite{jiang2018rational}. The computational approach presented here is time relevant and very important to understand the effects of the AD in cortical columns, but there is still work to be carried further. Since the relationship of neurons and astrocytes is not the focus of this paper, more efforts should appear in that regards by trying to incorporate astrocyte models in the cortical column models of the neurons such as the Digitalization of the Neocortical Column by the Blue Brain Project \cite{markram2006blue}. This computational effort can pave the way for a greater understanding of the multi-scale effects before the physical interfaces to start being implemented. Moreover, these bionano interfaces can probably be limited to some cell contact points and the study of computational approaches that can infer the detailed signal propagation inside cortical columns is very relevant, and the molecular communications community has a crucial role in that.

\begin{figure}
    \centering
    \includegraphics[width=\textwidth]{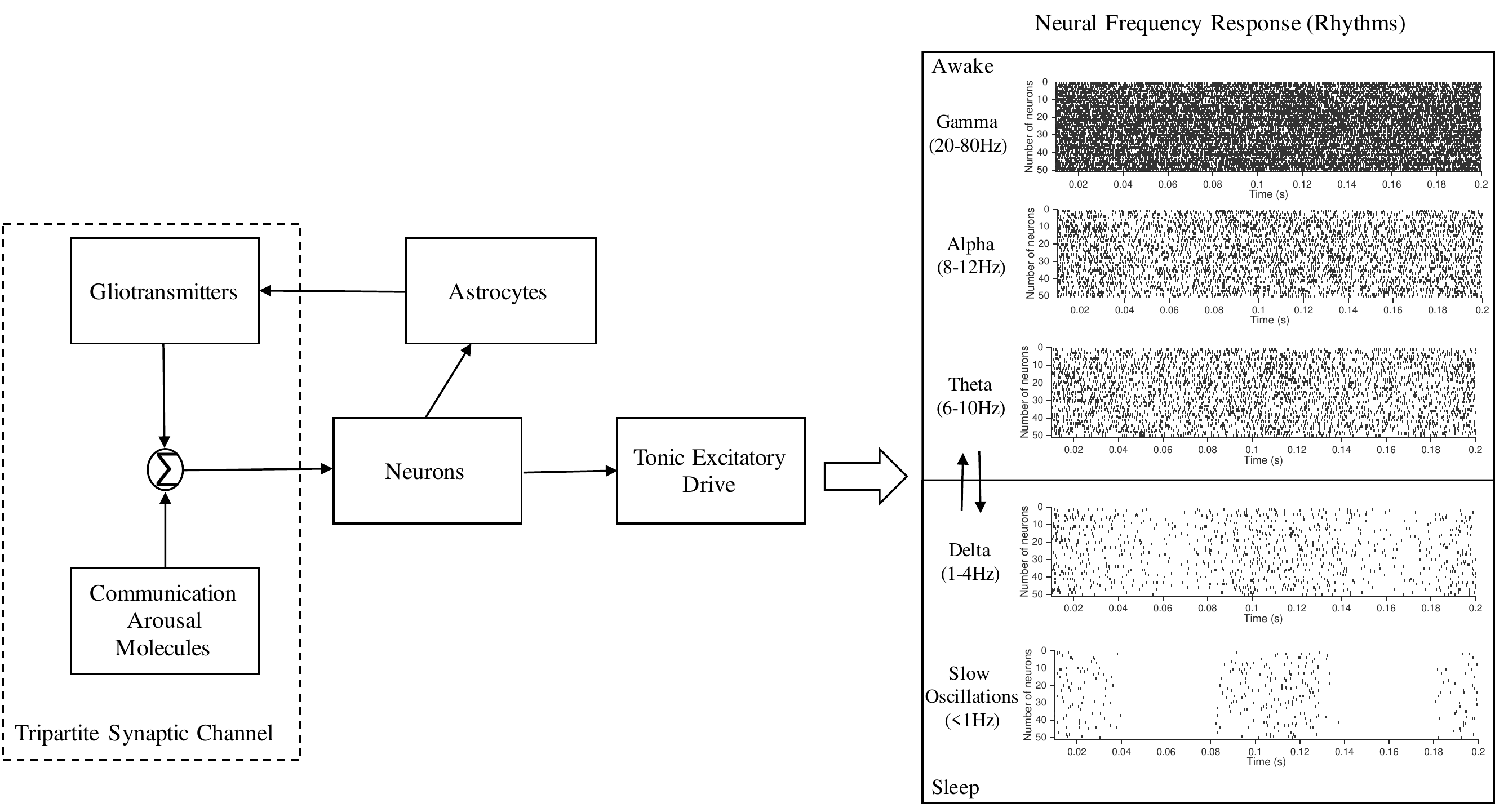}
    \caption{A descriptive model of the frequency response of neurons as an output of the tripartite synapses model. Both the giotransmitter and the communication arousal molecules are correlated to the neurons as inputs for the spike response of the neurons. The astrocytes will regulate this activity by regulating the gliotransmitters. The tonic excitatory drive converts the response of neurons into the brain waves or rhythms, between awake and sleep stages.}
    \label{fig:effect_neural_rythms}
\end{figure}

As briefly explained in the results, the neural rhythms or, in other words, the different frequencies values of spiking activity in neurons can change depending on the type of neurons and region of the brain. In Fig \ref{fig:effect_neural_rythms}, we present an initial model of the relationship of the tripartite synapses (connections between pre and postsynaptic neurons with the astrocytes through the synaptic cleft) and the neuron frequency response, inspired by \cite{berridge2016inositol}. This interesting relation between astrocytes and neurons is directly related to the neural rhythms \cite{lenk2016understanding,wallach2014glutamate}. The released gliotransmitter from astrocytes to the synaptic cleft has a significant role in the synaptic plasticity, which is the mechanism needed for ensuring the propagation of higher neuronal frequencies. The short-term memory is only formed through higher frequency neurons during the awake stage \cite{berridge2016inositol}. In the AD, the short-term memories are not formed if the synaptic plasticity is not strong enough  \cite{berridge2016inositol}. We showed in our paper that delay and gain, which relates to the time periods and calcium concentration in astrocytes, and it can potentially be linked to the weak short-term plasticity and loss of short-term memory. We also showed that astrocytes with AD will have an increased delay and decreased gain due to the influence of $\beta$-amyloid. That means this alteration of signal propagation performance over the networks could be a link to the frequency response and ultimately answer the effects of short-term memory loss from a network perspective. This is a very important approach for the early AD detection, and together with bionano sesing with new bionano interfaces, could find a way to bring hope to Alzheimer's patients and try to stop the progress of the disease by detecting it as early as possible.

% Another important aspect that directly affects the propagation of calcium in astrocytes network is the action potential propagation in neuronal networks.

% one paragraph on how to integrate two types of networks to create a increased realistic computation approach. the impact of it is a more carried out study of neural rhythms which could explain why, in alzheimer's, the short term memory will happen with a determined spatial propagation model of the $\beta$-amyloid.

% one paragraph about novel imaging and sensing techniques. talk about the difficult in detect early onset of alzheimer's, and talk about the important of computational models in this regard
% one paragraph about future treatment strategies

\section{Conclusions}

Future nanobio sensing techniques will rely upon inference techniques based on computational models of tissues to obtain a detailed description of cellular networks. In the brain, astrocyte networks present different types of topologies. They are severely impacted by the presence of $\beta$-amyloid plaques, which is linked as one of the causes of Alzheimer's disease. In this paper, we propose a computational approach that captures the effect of $\beta$-amyloid in the calcium signalling of astrocytes. It accounts for the changes in intracellular pathways, IP$_3$ dynamics, gap junctions, voltage-gated calcium channels and astrocytes volume. We implemented the shortcut networks, regular degree networks, Erd\"os R\'enyi networks and link radius networks topologies. Lastly, we designed a communication system to evaluate the propagation extend, molecular delay and channel gain metrics with frequency modulation of calcium signals. The results show that the more unstable but at the same time lower level oscillations of Alzheimer's astrocyte networks. This single cell effect can create a multi-scale impact on communication between astrocytes with the AD by changing the propagation extend in all network topologies, increasing the molecular delay and lowering the channel gain compared to healthy astrocytes. These observations are different in the many network topologies, but with an increased impact in Erd\"os R\'enyi networks and link radius networks topologies. Even though we have drafted a holistic approach to capturing the effect of the AD in astrocytes networks in a computational approach, more work needs to be carried further. For example, astrocytes networks with AD should be implemented with connections to neuron microcolumns of the cortex. Moreover, the effects of $\beta$-amyloid plaques need to be considered in the synaptic cleft and neurons for a more accurate characterisation of the impact of the AD in the brain.

\bibliographystyle{IEEEtran}
\bibliography{References}

\end{document}